\documentclass[prd,twocolumn,floatfix,nofootinbib]{revtex4}
\usepackage{amssymb}
\usepackage{amsmath}
\usepackage{graphicx,subfigure,color,dcolumn,booktabs,bm}
\usepackage{longtable,lscape}
\usepackage{txfonts}
\usepackage{overpic}
\usepackage{indentfirst}
\usepackage{cases}
\usepackage{multirow}
\usepackage{ulem}
\usepackage[colorlinks,
            citecolor=blue,
            anchorcolor=red,
            menucolor=red,
            linkcolor=red,
            filecolor=red,
            runcolor=red,
            urlcolor=blue,
            frenchlinks=false]{hyperref}

\graphicspath{{Figures/}}
\allowdisplaybreaks
\begin{document}

\title{Study of the hidden-heavy pentaquarks and $P_{cs}$ states}
\author{Wen-Xuan Zhang$^{1}$}
\email{zhangwx89@outlook.com}
\author{Chang-Le Liu$^{1}$}
\email{liuchanglelcl@qq.com}
\author{Duojie Jia$^{1,2}$\thanks{%
Corresponding author}}
\email{jiadj@nwnu.edu.cn}
\affiliation{$^1$Institute of Theoretical Physics, College of Physics and Electronic
Engineering, Northwest Normal University, Lanzhou 730070, China \\
$^2$Lanzhou Center for Theoretical Physics, Lanzhou University, Lanzhou,
730000, China \\
}
\date{\today}

\begin{abstract}
In light of the recently observed resonance states $P_{\psi s}^{\Lambda}(4338)^0$ and $P_{cs}(4459)^0$ by LHCb Collaboration
in $J/\psi\Lambda$ decay channel, we perform a systematical study of all possible hidden-heavy pentaquarks with strangeness $S=0,-1,-2,-3$, 
in unified framework of MIT bag model. The color-spin wavefunctions presented in terms of Young-Yamanouchi bases and transformed into baryon-meson couplings, 
are utilized to calculate masses, magnetic moments and ratios of partial widths. With numerical analysis, the observed $P_{\psi s}^{\Lambda}(4338)^0$ is likely 
to be a $1/2^-$ compact $P_{cs}$ pentaquark, and $P_{cs}(4459)^0$ favors two-peak structure of $3/2^-$ and $1/2^-$ $P_{cs}$ states. Further predictions
on hadron properties and decay channels are given to compact $P_{css}$, $P_{csss}$ states and bottom sectors.

PACS number(s):12.39Jh, 12.40.Yx, 12.40.Nn

\end{abstract}

\maketitle
\date{\today}

\section{Introduction}
\label{sec:intro}

In addition to the conventional hadronic states, such as mesons $q\bar{q}$ and baryons $qqq$ in quark configurations, 
the possible exotic tetraquarks $q^2\bar{q}^2$ and pentaquarks $q^4\bar{q}$ are also suggested at the birth of quark model \cite{Gell-Mann:1964ewy,Zweig:1964ruk}.
Later in the 1970s, the MIT bag model has been developed by Jaffe \cite{Jaffe:1976ig,Jaffe:1976ih,DeGrand:1975cf} for the study of 
exotic multiquark states and notation of color confinement. In the past few decades, it has been applied to describe the doubly heavy baryons \cite{Fleck:1989mb,He:2004px,Bernotas:2008bu,Bernotas:2012nz}
and exotic states, including light pentaquarks \cite{Strottman:1979qu} and hybrid mesons \cite{Barnes:1982tx,Chanowitz:1982qj}.
Despite that these states are considered to be exotic beyond the conventional scheme of quark model, the existences of them are allowed by quantum chromodynamics (QCD).

Since the observation of exotic $X(3872)$ in 2003 by the Belle \cite{Belle:2003nnu}, there are many candidates that have been discovered for tetraquarks,
such as the $Z_c(3900)$ \cite{BESIII:2013ris,Belle:2013yex} and the $T_{cc}(3875)$ \cite{LHCb:2021auc}, as well as the fully charm systems $X(6600)$ \cite{CMS:2023owd} 
and $X(6900)$ \cite{LHCb:2020bwg}. In 2015, the first evidence for pentaquark-like structures $P_c(4450)^+$ and $P_c(4380)^+$ with a minimal quark constituent of
$uudc\bar{c}$ was reported by LHCb in $J/\psi p$ channel \cite{LHCb:2015yax}, for which the former exhibits a two-peak structure resolved into $P_c(4440)^+$ and
$P_c(4457)^+$ in 2019. Recently, the LHCb Collaboration has reported two new hidden charm pentaquarks with a single strange flavor, $P_{\psi s}^{\Lambda}(4338)^0$ \cite{LHCb:2022ogu},
and $P_{cs}(4459)^0$\cite{LHCb:2020jpq} both in $J/\psi\Lambda$ channel. These pentaquark candidates encourage the theoretical study on their mass spectrum,
hadron properties and decay behaviors. There are various pictures and methods applied to analyze the hidden-charm pentaquarks, including molecular  
\cite{Chen:2020uif,Chen:2020kco,Yang:2021pio,Lu:2021irg,Xiao:2021rgp,Zhu:2021lhd,Feijoo:2022rxf,Zhu:2022wpi,Chen:2022wkh,Wang:2022mxy,Wang:2023iox,Azizi:2023iym} and compact scenario
\cite{Weng:2019ynv,Shi:2021wyt,Ruangyoo:2021aoi,Li:2021ryu,Li:2023aui,Guo:2023fih}, as well as the hidden-bottom pentaquarks \cite{Yang:2018oqd,Zhu:2020vto}.

In our previous work \cite{Zhang:2021yul}, we performed a systematical study on the spectrum of mesons and baryons in their ground states, 
including light-flavor baryons and doubly heavy baryon $\Xi_{cc}$ \cite{LHCb:2017iph}, and extented the computation to the doubly heavy tetraquark $T_{cc}$.
Considering the binding energies between heavy quarks, along side a running coupling constant, our predictions align with the experimental reports
for baryons and tetraquarks with two heavy flavors, even for the fully heavy system $X(6600)$ \cite{Yan:2023lvm}.
Therefore, the MIT bag model serves as a capable method to study heavy hadrons on masses and magnetic moments.
\begin{figure}[t]
    \centering
    \includegraphics[width=0.45\textwidth]{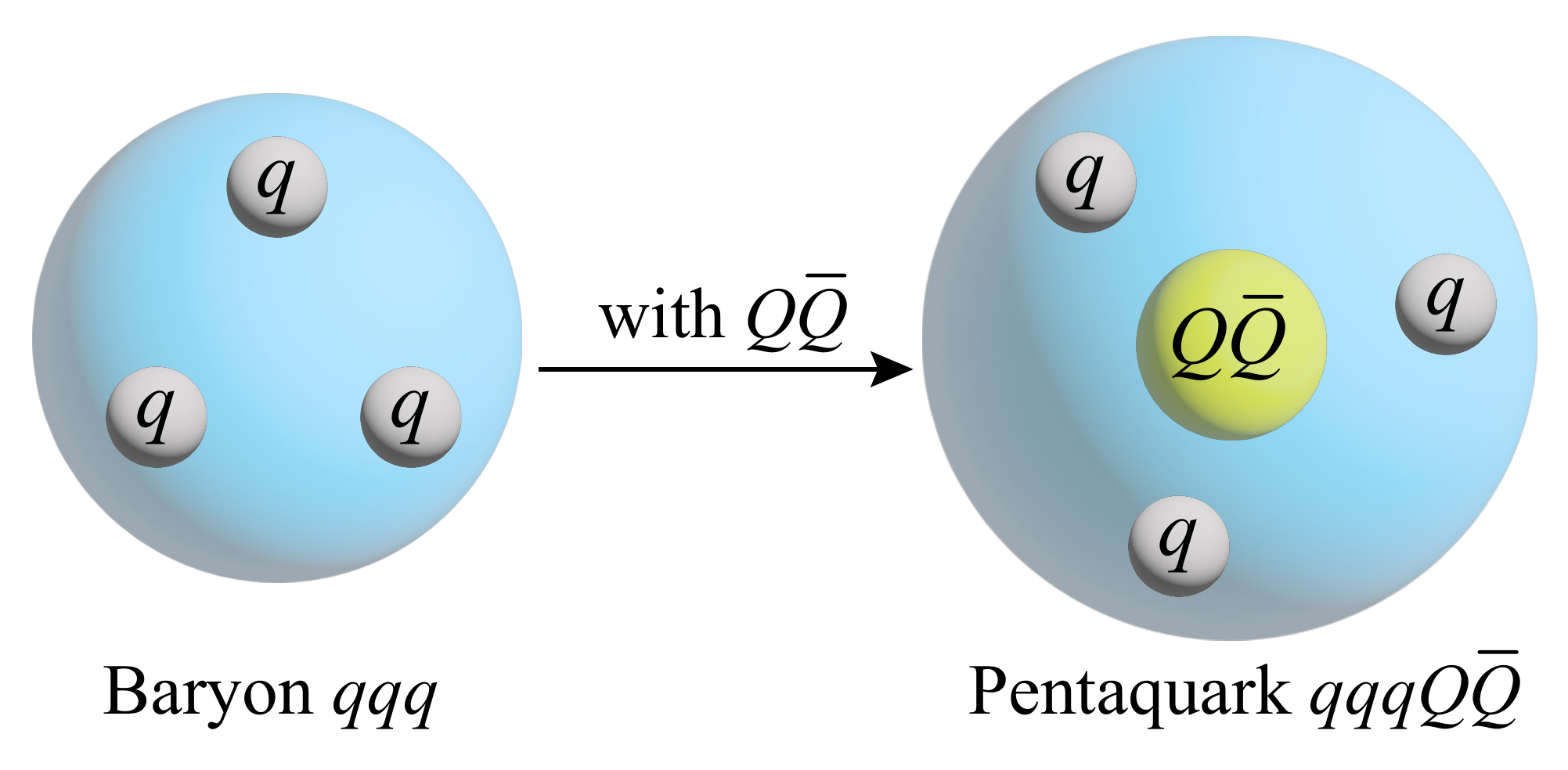}
    \caption{The relationship of inner-structures between light-flavor baryon and hidden-heavy pentaquark. }
    \label{fig:qqq-qqqQQ}
\end{figure}
Motivated by this, as demonstrated in Fig.(\ref{fig:qqq-qqqQQ}), a light-flavor baryon $qqq$ ($q=u,d,s$) described as a sphere bag
is extended into a hidden-heavy pentaquark $qqqQ\bar{Q}$, with the present of heavy $Q\bar{Q}$ ($Q=c,b$) in configuration.
Due to the dynamics of heavy constituents and suppression of relativistic effects, it is possible to consider
pentaquarks $qqqQ\bar{Q}$ in compact scenario.

The purpose of this work is to study masses, magnetic moments and ratios of partial widths for hidden-heavy pentaquarks,
classified into strangeness $S=0,-1,-2,-3$, namely $nnnQ\bar{Q}$, $nnsQ\bar{Q}$, $ssnQ\bar{Q}$ and $sssQ\bar{Q}$, respectively.
Numerical computations are based on the unified framework of MIT bag model, characterized by model parameters and fundamental relations.
In order for describing inner-structures of pentaquarks, the color-spin wavefunctions are employed to the chromomagnetic interactions,
which are expressed in terms of Young tableau and Young-Yamanouchi bases. For partial width study, we transform the color-spin wavefunctions
into the form of color-singlet (baryon-meson coupling) and color-octet components. Finally, with the help of spin bases, 
it's straightforward to derive the magnetic moments for possible pentaquark configurations.

This work is structured as follows. In the next section, we provide a brief introduction to the fundamental relations 
of the Hamiltonian and magnetic moment. Additionally, we outline the parameters utilized in our numerical calculations. 
The detailed formulations about the chromomagnetic structure and magnetic moment can be found in the Appendix.
In Section \ref{sec:result}, we present a comprehensive numerical analysis of hidden-heavy pentaquarks, classified into subsections
\ref{sec:nnnQQ}, \ref{sec:nnsQQ}, \ref{sec:ssnQQ}, and \ref{sec:sssQQ}, corresponding to strangeness $S=0,-1,-2,-3$, respectively.
Finally, this work ends with summaries and conclusions in Section \ref{sec:summary}.

\section{Mass and magnetic moment}
\label{sec:mothed}
In the framework of the MIT bag model, all valence quarks are confined within a spherical bag characterized by a radius $R$ and respective momentum $x_i$. 
These quarks are coupled with perturbative chromomagnetic interactions facilitated by the exchange of the lowest-order gluon, 
as described in prior works \cite{DeGrand:1975cf,Johnson:1975zp}. The mass formula associated with the bag, representing a hadronic state, is given by
\begin{equation}
    M\left( R\right) =\sum_{i}\omega_{i}+\frac{4}{3}\pi R^{3}B-\frac{Z_{0}}{R}+M_{BD}+M_{CMI}, \label{equ:mass}
\end{equation}
\begin{equation}
    \omega_{i} =\left( m_{i}^{2}+\frac{x_{i}^{2}}{R^{2}}\right) ^{1/2}. \label{equ:omega}
\end{equation}
Here, in Eq.(\ref{equ:mass}), the first term represents the sum of the kinetic energy for the relativistic quark $i$ with mass $m_{i}$,
the second is volume energy with bag constant $B$, which characterizes the difference between perturbative and non-perturbative QCD vacuum,
the third is zero point energy with coefficient $Z_{0}$, 
the forth accounts for the binding energy between heavy quarks or heavy and strange quarks \cite{Karliner:2014gca,Karliner:2017elp,Karliner:2017qjm,Karliner:2020vsi},
and the fifth  represents the chromomagnetic interaction \cite{DeRujula:1975qlm}.
The bag radius $R$ is the only variable to be determined by minimizing Eq.(\ref{equ:mass}).
The momentum $x_{i}$, given in units of $R^{-1}$ satisfies a boundary condition on the bag surface
\begin{equation}
    \tan x_{i}=\frac{x_{i}}{1-m_{i}R-\left( m_{i}^{2}R^{2}+x_{i}^{2}\right)
    ^{1/2}}, \label{equ:boundary}
\end{equation}
which can be derived by quark spinor wave function.

The interactions among quarks involved in this work have two primary components: 
the lowest-order gluon exchange and short-range effects \cite{DeRujula:1975qlm}. 
For the first part, the binding energies $M_{BD}$ primarily arise from short-range chromoelectric interactions 
between heavy quarks or heavy-strange systems, both of which are massive and moving nonrelativistically.
Specifically, we incorporate four binding energies $B_{cs}$, $B_{cc}$, $B_{bs}$ and $B_{bb}$ 
in color $\boldsymbol{\bar{3}}_{c}$ representation for hidden-heavy pentaquarks, 
and scale them to other color configurations by color factors \cite{Karliner:2014gca,Karliner:2017qjm}.
The second component involves chromomagnetic interaction $M_{CMI}$, which contributes to the overall dynamics of the system.
The interaction arising from the magnetic moments of quark spin can be described by the following typical form:
\begin{equation}
    M_{CMI}=-\sum_{i<j}\left\langle \boldsymbol{\lambda _{i}}\cdot \boldsymbol{\lambda
    _{j}}\right\rangle \left\langle \boldsymbol{\sigma _{i}}\cdot \boldsymbol{\sigma _{j}}
    \right\rangle C_{ij}, \label{equ:CMI}
\end{equation}
where $i$ and $j$ represent the indices of the quarks (or anti-quarks), $\lambda$ denotes the Gell-Mann matrices, 
$\sigma$ signifies the Pauli matrices, and $C_{ij}$ corresponds to the chromomagnetic interaction (CMI) parameters as defined in Ref. \cite{Zhang:2021yul}. 
The color and spin factor (average value of Casimir operator) 
in Eq.(\ref{equ:CMI}) can be calculated utilizing the specific method in Refs.\cite{Zhang:2021yul,Zhang:2023hmg,Yan:2023lvm},
with the color-spin wavefunctions given in Appendix A.

Next, we will provide a brief overview of the foundational equations of magnetic moment.
In the context of MIT bag model, the magnetic moment for an individual quark is not determined by constant parameters
but rather by variables that satisfy specific boundary conditions. Following variational calculations on mass equation (\ref{equ:mass}), 
the model fulfills the quark spinor wavefunction \cite{DeGrand:1975cf}
\begin{equation}
    \psi_{i}(r) = N_{i} \binom{j_{0}(x_{i}r/R)U}
    {i\frac{x_{i}}{(\omega_{i}+m_{i})R}j_{1}(x_{i}r/R)\boldsymbol{\sigma}\cdot\boldsymbol{\hat{r}}U} e^{-i\omega_{i}t}, \label{equ:quarkspinor}
\end{equation}
with parameters $(R,x_i)$ and mass $m_i$ for quark $i$, which would help determine the physical quantities and interactions within the bag model.
Subsequently, the operator $\boldsymbol{r\times\gamma}$ is applied to the wavefunction (\ref{equ:quarkspinor}) to derive the magnetic moment, as the following expression:
\begin{equation}
    \begin{aligned}
        \mu_{i} &= \frac{Q_{i}}{2} \int_{bag}\mathrm{d}^{3}r\, 
        \bar{\psi_{i}}\left(\boldsymbol{r\times\gamma}\right)\psi_{i} \\
        &= \frac{Q_{i}}{2} \int_{0}^{R}\mathrm{d}r\,r^{2} \int\mathrm{d}\Omega\
        \bar{\psi_{i}}\left(\boldsymbol{r\times\gamma}\right)\psi_{i} \\
        &= Q_{i}\frac{R}{6}\frac{4\omega_{i}R+2m_{i}R-3}{2\omega_{i}R\left(\omega_{i}R-1\right)+m_{i}R},
    \end{aligned}\label{equ:mui}
\end{equation}
where $Q_i$ is electric charge of quark and $\boldsymbol{\gamma}$ are Dirac matrices for spinor fields.
Accordingly, it's straightforward to apply Eq.(\ref{equ:mui}) to calculate the magnetic moment for any flavor of quark 
with mass $m_i$ and parameters $(R,x_i)$ evaluated through spectrum studies.

We calculate the magnetic moment of hadron by summing up Eq.(\ref{equ:mui}) via color-spin wavefunction $\left|\psi\right\rangle$, 
resulting in the expression \cite{Wang:2016dzu}
\begin{equation}
    \mu =\left\langle \psi \left\vert\hat{\mu}\right\vert \psi \right\rangle, \quad
    \hat{\mu} = \sum\nolimits_{i}g_{i}\mu_{i}\hat{S}_{iz}, \label{equ:musum}
\end{equation}
with $g_{i}=2$ and $\hat{S}_{iz}$ denotes the third component of spin for an individual quark.
The numerical results will be related to the magnetic moment of proton, and transformed into that in unit of $\mu_{N}$, 
with the help of measured data $\mu_{p}=2.79285\mu_{N}$ \cite{ParticleDataGroup:2022pth,Tiesinga:2021myr}.
Eq.(\ref{equ:musum}) holds true for chromomagnetic mixing, as the spin wavefunctions expanding to basis vectors create matrix elements of magnetic moment.
For further details, refer to Appendix B.

In numerical calculations, we proceed with model parameters from a previous work \cite{Zhang:2021yul},
that successfully reconcile ground-state mesons and baryons and provide descriptions of fully heavy tetraquarks and pentaquarks \cite{Zhang:2023hmg,Yan:2023lvm}.
The chosen constants and masses for quarks with flavor $n$ (light nonstrange flavor $n=u,d$), 
strange $s$, charm $c$ and bottom $b$ are as follows:
\begin{equation}
    \begin{Bmatrix}
        Z_{0}=1.83, & B^{1/4}=0.145\,\text{GeV}, \\
        m_{n}=0\,\text{GeV}, & m_{s}=0.279\,\text{GeV}, \\
        m_{c}=1.641\,\text{GeV}, & m_{b}=5.093\,\text{GeV}.
    \end{Bmatrix}\label{equ:parameter}
\end{equation}
Meanwhile, the binding energies for color $\boldsymbol{\bar{3}}_{c}$ representation are considered:
\begin{equation}
    \begin{Bmatrix}
        B_{cs}=-0.025\,\text{GeV}, & B_{cc}=-0.077\,\text{GeV}, \\
        B_{bs}=-0.032\,\text{GeV}, & B_{bb}=-0.128\,\text{GeV}.
    \end{Bmatrix}\label{equ:binding}
\end{equation}
These bindings will be scaled by color factors.
With the parameters in Eqs.(\ref{equ:parameter}) and (\ref{equ:binding}),
the variational method can be performed to determine the bag radius $R$ and respective momentum $x_{i}$
using Eq.(\ref{equ:boundary}) for pentaquarks with quantum numbers $IJ^P$.
Subsequently, the parameters $(R,x_i)$ enable the calculations of masses and magnetic moments for the hidden-heavy pentaquarks.

\section{Analysis of spectrum and decay}
\label{sec:result}

\subsection{The $nnnQ\bar{Q}$ systems}
\label{sec:nnnQQ}

\renewcommand{\tabcolsep}{0.12cm}
\renewcommand{\arraystretch}{1.2}
\begin{table*}[!htbp]
    \caption{Calculated spectra (in GeV) of pentaquarks $nnnb\bar{b}$. Bag radius $R_0$ is in GeV$^{-1}$. 
    Magnetic moments are in unit of $\mu_N$, and organized in the order of $I_3=3/2, 1/2, -1/2, -3/2$ for $I=3/2$, or $I_3=1/2, -1/2$ for $I=1/2$.
    The numbers below respective decay channels are ratios of partial width.
    The states denoted by asterisks couple strongly to scattering states.}
    \label{tab:nnnbb}
    \begin{tabular}{ccccc|cccc|cccccc}
        \bottomrule[1.5pt]\bottomrule[0.5pt]
        \multirow{2}{*}{$I$} &\multirow{2}{*}{$J^{P}$}
        &\multicolumn{3}{l|}{$nnnb\bar{b}$}
        &\multicolumn{4}{l|}{$nnn\otimes b\bar{b}$} &\multicolumn{6}{l}{$nnb\otimes n\bar{b}$} \\
        & &$R_{0}$ &$M$ &$\mu$ &$\Delta\Upsilon$ &$\Delta\eta_{b}$ &$N\Upsilon$ &$N\eta_{b}$
        &$\Sigma_{b}^{\ast}B^{\ast}$ &$\Sigma_{b}^{\ast}B$ &$\Sigma_{b}B^{\ast}$ &$\Sigma_{b}B$
        &$\Lambda_{b}B^{\ast}$ &$\Lambda_{b}B$ \\ \hline
        3/2
            &${5/2}^{-}$    &5.52   &11.235  & &$\ast$ & & & & & & & & & \\
            &${3/2}^{-}$    &5.53   &11.561  &1.99, 0.99, 0.00, -0.99 &0 &1 & & &1 &0.58 &0.18 & & & \\
            &               &5.52   &11.235  & &$\ast$ & & & & & & & & & \\
            &               &5.50   &11.230  & & &$\ast$ & & & & & & & & \\
            &${1/2}^{-}$    &5.57   &11.583  &0.64, 0.27, -0.10, -0.47 &1 & & & &1 & &0.15 &0.03 & & \\
            &               &5.50   &11.539  &0.68, 0.39, 0.10, -0.19 &1 & & & &1 & &13.43 &5.92 & & \\
            &               &5.52   &11.233  & &$\ast$ & & & & & & & & & \\
        1/2
            &${5/2}^{-}$    &5.52   &11.431  &2.96, 0.00 & & & & &1 & & & & & \\
            &${3/2}^{-}$    &5.50   &11.412  &2.78, -0.04 & & &1 & &2.48 &1 &0.66 & &1 & \\
            &               &5.46   &11.394  &2.18, 0.05 & & &1 & &0.07 &1 &3.25 & &1 & \\
            &               &5.45   &11.333  &1.09, -0.01 & & &1 & &0.27 &1 &0.09 & &1 & \\
            &               &5.43   &10.929  & & & &$\ast$ & & & & & & & \\
            &${1/2}^{-}$    &5.46   &11.380  &1.49, 0.01 & & &1 &3.64 &1 & &1.23 &1.31 &1 &0.08 \\
            &               &5.43   &11.323  &0.80, -0.06 & & &1 &0.0005 &1 & &0.22 &1.23 &1 &1.81 \\
            &               &5.42   &11.314  &-0.03, 0.06 & & &1 &14.66 &1 & &4.78 &5.36 &1 &0.77 \\
            &               &5.42   &10.929  & & & &$\ast$ & & & & & & & \\
            &               &5.41   &10.923  & & & & &$\ast$ & & & & & & \\
        \bottomrule[0.5pt]\bottomrule[1.5pt]
    \end{tabular}
\end{table*}

\renewcommand{\tabcolsep}{0.12cm}
\renewcommand{\arraystretch}{1.2}
\begin{table*}[!htbp]
    \caption{Calculated spectra (in GeV) of pentaquarks $nnnc\bar{c}$. Bag radius $R_0$ is in GeV$^{-1}$. 
    Magnetic moments are in unit of $\mu_N$, and organized in the order of $I_3=3/2, 1/2, -1/2, -3/2$ for $I=3/2$, or $I_3=1/2, -1/2$ for $I=1/2$.
    The numbers below respective decay channels are ratios of partial width.
    The states denoted by asterisks couple strongly to scattering states.}
    \label{tab:nnncc}
    \begin{tabular}{ccccc|cccc|cccccc}
        \bottomrule[1.5pt]\bottomrule[0.5pt]
        \multirow{2}{*}{$I$} &\multirow{2}{*}{$J^{P}$}
        &\multicolumn{3}{l|}{$nnnc\bar{c}$}
        &\multicolumn{4}{l|}{$nnn\otimes c\bar{c}$} &\multicolumn{6}{l}{$nnc\otimes n\bar{c}$} \\
        & &$R_{0}$ &$M$ &$\mu$ &$\Delta J/\psi$ &$\Delta\eta_{c}$ &$N J/\psi$ &$N\eta_{c}$
        &$\Sigma_{c}^{\ast}D^{\ast}$ &$\Sigma_{c}^{\ast}D$ &$\Sigma_{c}D^{\ast}$ &$\Sigma_{c}D$
        &$\Lambda_{c}D^{\ast}$ &$\Lambda_{c}D$ \\ \hline
        3/2
            &${5/2}^{-}$    &5.81   &4.547 & &$\ast$ & & & & & & & & & \\
            &${3/2}^{-}$    &5.82   &4.758 &2.23, 1.12, 0.00, -1.12 &0 &1 & & &1 &0.53 &0.13 & & & \\
            &               &5.81   &4.547 & &$\ast$ & & & & & & & & & \\
            &               &5.71   &4.503 & & &$\ast$ & & & & & & & & \\
            &${1/2}^{-}$    &5.91   &4.820 &1.59, 1.07, 0.56, 0.04 &1 & & & &1 & &0.09 &0.02 & & \\
            &               &5.76   &4.703 &-0.01, -0.27, -0.52, -0.78 &1 & & & &1 & &36.85 &10.1 & & \\
            &               &5.75   &4.524 & &$\ast$ & & & & & & & & & \\
        1/2
            &${5/2}^{-}$    &5.85   &4.661 &3.13, 0.00 & & & & &1 & & & & & \\
            &${3/2}^{-}$    &5.83   &4.630 &2.85, 0.29 & & &1 & &4.13 &1 &0.26 & &1 & \\
            &               &5.74   &4.569 &1.69, -0.55 & & &1 & &0.02 &1 &19.36 & &1 & \\
            &               &5.68   &4.503 &1.83, 0.27 & & &1 & &- &1 &0.56 & &1 & \\
            &               &5.73   &4.241 & & & &$\ast$ & & & & & & & \\
            &${1/2}^{-}$    &5.80   &4.580 &1.18, -0.08 & & &1 &0.39 &1.32 & &1 &0.25 &1 &0.14 \\
            &               &5.70   &4.490 &0.06, 0.25 & & &1 &10.99 &- & &1 &0.01 &1 &9.3 \\
            &               &5.61   &4.452 &1.12, -0.18 & & &1 &2.73 &- & &- &1 &1 &0.004 \\
            &               &5.71   &4.235 & & & &$\ast$ & & & & & & & \\
            &               &5.59   &4.191 & & & & &$\ast$ & & & & & & \\
        \bottomrule[0.5pt]\bottomrule[1.5pt]
    \end{tabular}
\end{table*}

In the initial phase of this study, we focus on the $nnnQ\bar{Q}$ ($Q=c$,$b$) pentaquark states, with mass spectrum, magnetic moments and partial width analysis.
Spectrum calculations are conducted employing the variational method of MIT bag model, accounting for chromomagnetic interactions.
To explore partial width and stability, we apply the bases in Ref.\cite{Weng:2019ynv} to probe the baryon-meson coupling components within the wavefunctions.
The numerical results are tabulated in Tables \ref{tab:nnnbb} and \ref{tab:nnncc}.

Before delving into discussions on hadron properties, it's important to exclude scattering states.
The chromomagnetic interaction allows pentaquarks $nnnQ\bar{Q}$ to exhibit three to five eigenstates.
However, some of them are loosely bound and inherently unstable, with a very broad decay width.
In order to filter such states, we examine the color-spin wavefunction $\left|\psi\right\rangle$ using the corresponding bases and eigenvectors obtained in this work.
We express $\left|\psi\right\rangle$ in terms of color-singlet $\mathbf{1_c}$ and color-octet $\mathbf{8_c}$ as follows:
\begin{equation}
    \left|\psi\right\rangle = c_{1} \left|q_{1}q_{2}q_{3}\right\rangle^{\mathbf{1}}_{S_1} \left|q_{4}\bar{q_{5}}\right\rangle^{\mathbf{1}}_{S_2} +
    c_{2} \left|q_{1}q_{2}q_{3}\right\rangle^{\mathbf{8}}_{S_3} \left|q_{4}\bar{q_{5}}\right\rangle^{\mathbf{8}}_{S_4} + \dots. \label{equ:component}
\end{equation}
In this expression, the first component represents the dissociation into $S$-wave baryon and meson directly (with spin $S_1$ and $S_2$),
and the coefficient $c_1$ is the overlap of the wavefunction calculated by diagonalization of chromomagnetic interaction matrix.
If a pentaquark couples strongly to a scattering state, signifying that the probability ${|c_1|}^2$ approaches 1, we will disregard and denote it 
with an asterisk in Tables \ref{tab:nnnbb} and \ref{tab:nnncc}, for the corresponding decay channel.

Apart from stability research, the eigenvectors play a crucial role in investigating the ratio of partial width for the respective decay channel.
Drawing from previous studies \cite{Weng:2019ynv,Weng:2020jao,Weng:2021ngd,An:2020vku}, 
we employ the formula for two body $L$-wave decay \cite{gaoc1992}:
\begin{equation}
    \Gamma_i = \gamma_i\alpha \frac{k^{2L+1}}{m^{2L}}\cdot {|c_i|}^2. \label{equ:partialwidth}
\end{equation}
Here, $\Gamma_i$ represents the partial width of channel $i$, $\gamma_i$ is associated to decay dynamics, $\alpha$ signifies the coupling constant, 
$m$ stands for the mass of initial state, $k$ represents the momentum of final states in the rest frame, and $c_i$ denotes the coefficient of corresponding component.
For $S$-wave OZI-superallowed decay mode \cite{Jaffe:1976ig,Strottman:1979qu} in this work, 
the factors reduce to $\gamma_i\alpha k\cdot {|c_i|}^2$, where $k$ satisfies the equation 
\begin{equation}
    m_A=\sqrt{m_B^2+k^2}+\sqrt{m_C^2+k^2},
\end{equation}
for the process $A\to B+C$, and $c_i$ can be extracted from Eq.(\ref{equ:component}) as $c_1$.
The coefficient $\gamma_i$ depending on spatial wavefunctions of initial and final states,
remains consistent between vector and scalar mesons. Additionally, in the heavy quark limit, 
$\gamma_i$ also remains the same between $\Sigma_c^{\ast}$ and $\Sigma_c$ \cite{Weng:2021ngd}.
Therefore, the following relations hold:
\begin{equation}
    \begin{aligned}
        &\gamma_{\Delta\Upsilon} = \gamma_{\Delta\eta_b}, \quad \gamma_{N\Upsilon} = \gamma_{N\eta_b}, \\
        &\gamma_{\Delta J/\psi} = \gamma_{\Delta\eta_c}, \quad \gamma_{N J/\psi} = \gamma_{N\eta_c}, \\
        &\gamma_{\Sigma_b^{\ast}B^{\ast}} = \gamma_{\Sigma_b^{\ast}B} = \gamma_{\Sigma_b B^{\ast}} = \gamma_{\Sigma_b B}, \\
        &\gamma_{\Sigma_c^{\ast}D^{\ast}} = \gamma_{\Sigma_c^{\ast}D} = \gamma_{\Sigma_c D^{\ast}} = \gamma_{\Sigma_c D}, \\
        &\gamma_{\Lambda_b B^{\ast}} = \gamma_{\Lambda_b B}, \quad \gamma_{\Lambda_c D^{\ast}} = \gamma_{\Lambda_c D},
    \end{aligned}
\end{equation}
These relations enable the study of ratios of partial widths from factors $k\cdot {|c_i|}^2$.
The corresponding results are shown in Tables \ref{tab:nnnbb} and \ref{tab:nnncc} below the respective decay channels.
Forbidden processes due to mass conservation are denoted by a short-dash.

Upon the completion of numerical computations, we proceed to discuss the mass spectra and label a $nnnQ\bar{Q}$ pentaquark state into symbol $P_{Q}(I,J^P,M)$.
The scattering states exhibiting the lowest mass splittings, such as $P_{b}(1/2,1/2^-,10.923)$ and $P_{c}(1/2,1/2^-,4.191)$,
restrict the mass range for $nnnb\bar{b}$ to 11.31-11.58$\,$GeV and for $nnnc\bar{c}$ to 4.45-4.82$\,$GeV.
Obviously, the mass gap of 270$\,$MeV in $b$-sector is narrower than the 370$\,$MeV in $c$-sector due to the suppression of heavy quark. 
However, experimental reports do not fall within our computed mass range.
For states that may carry negative parity, there are
\begin{equation}
    \begin{aligned}
        &P_c(4312)^+ \ M=4312\,\textrm{MeV} \ \Gamma=9.8\,\textrm{MeV} \text{\cite{LHCb:2019kea}}, \\
        &P_c(4337)^+ \ M=4337\,\textrm{MeV} \ \Gamma=29\,\textrm{MeV} \text{\cite{LHCb:2021chn}}, \\
        &P_c(4380)^+ \ M=4380\,\textrm{MeV} \ \Gamma=215\,\textrm{MeV} \text{\cite{LHCb:2015yax}},
    \end{aligned}
\end{equation}
all below the state $P_{c}(1/2,1/2^-,4.452)$ in mass.
This suggests that they are unlikely to be compact pentaquarks, a conclusion supported by molecular models \cite{Burns:2021jlu,Ling:2021lmq,Du:2021fmf,Wang:2021crr,Chen:2022onm}.

Nevertheless, several states appear to be potentially compact, including the lightest $P_{b}(1/2,1/2^-,11.380)$, $P_{b}(1/2,1/2^-,11.323)$ 
and $P_{b}(1/2,1/2^-,11.314)$ for $b$ system and $P_{c}(1/2,1/2^-,4.580)$, $P_{c}(1/2,1/2^-,4.490)$ and $P_{c}(1/2,1/2^-,4.452)$ for $c$ system.
In the $nnn\otimes Q\bar{Q}$ coupling, we determine the ratios of partial widths:
\begin{equation}
    \begin{aligned}
        \frac{\Gamma(P_{c}(1/2,1/2^-,4.580) \to N\eta_c)}{\Gamma(P_{c}(1/2,1/2^-,4.580) \to NJ/\psi)} &= 0.39, \\
        \frac{\Gamma(P_{c}(1/2,1/2^-,4.490) \to N\eta_c)}{\Gamma(P_{c}(1/2,1/2^-,4.490) \to NJ/\psi)} &= 10.99, \\
        \frac{\Gamma(P_{c}(1/2,1/2^-,4.452) \to N\eta_c)}{\Gamma(P_{c}(1/2,1/2^-,4.452) \to NJ/\psi)} &= 2.73,
    \end{aligned}
\end{equation}
which could potentially predict the decay into $N\eta_c$ channel.
Notably, some states exhibit dominant channels, such as $P_{b}(1/2,1/2^-,11.323)$ decaying into $N\Upsilon$
or the one with $(I)J^P=(3/2)3/2^-$ decaying into $\Delta\eta_c$ or $\Delta\eta_b$.
If experiments report resonances with masses close to our predictions but in the opposite decay channel, 
this might either exclude them from our predicted spectrum or explain them as compact pentaquarks.
In the $nnQ\otimes n\bar{Q}$ coupling, the decay behaviors are studied, awaiting confirmation through experimental findings.
Particularly, the state $P_{c}(1/2,1/2^-,4.452)$ with limited decay channels $\Sigma_c D$ and $\Lambda_c D^\ast$
can be discoverable in corresponding processes.

In the fifth column of Tables \ref{tab:nnnbb} and \ref{tab:nnncc}, we present the magnetic moments of pentaquarks
while excluding scattering states. Similar evaluations have been conducted in Refs.\cite{Wang:2016dzu,Li:2021ryu,Guo:2023fih}
covering various configurations. However, this work does not delve into transition moments as $M1$ transition processes are typically suppressed against strong decay.
The magnetic moments are organized based on the isospin component $I_3$, ranging from $0.09\mu_N$ to $3.35\mu_N$.
This implies that the electric charge of the final states could potentially aid in predicting the specific magnetic moment, 
if the mass and decay channel align with experimental reports.

\subsection{The $nnsQ\bar{Q}$ systems}
\label{sec:nnsQQ}

\renewcommand{\tabcolsep}{0.12cm}
\renewcommand{\arraystretch}{1.0}
\begin{table*}[!htbp]
    \caption{Calculated spectra (in GeV) of pentaquarks $nnsb\bar{b}$. 
    Magnetic moments are in unit of $\mu_N$, and organized in the order of $I_3=1, 0, -1$ for $I=1$, or $I_3=0$ for $I=0$.
    The bag radius $R_0$ is determined to be 5.50$\,$GeV$^{-1}$.
    The numbers below respective decay channels are ratios of partial width.
    The states denoted by asterisks couple strongly to scattering states.}
    \label{tab:nnsbb}
    \begin{tabular}{cccc|cccccc|cccccc}
        \bottomrule[1.5pt]\bottomrule[0.5pt]
        \multirow{2}{*}{$I$} &\multirow{2}{*}{$J^{P}$}
        &\multicolumn{2}{l|}{$nnsb\bar{b}$}
        &\multicolumn{6}{l|}{$nns\otimes b\bar{b}$} &\multicolumn{6}{l}{$nnb\otimes s\bar{b}$} \\
        & &$M$ &$\mu$ &$\Sigma^{\ast}\Upsilon$ &$\Sigma^{\ast}\eta_{b}$
        &$\Sigma\Upsilon$ &$\Sigma\eta_{b}$ &$\Lambda\Upsilon$ &$\Lambda\eta_{b}$
        &$\Sigma_{b}^{\ast}B_{s}^{\ast}$ &$\Sigma_{b}^{\ast}B_{s}$ &$\Sigma_{b}B_{s}^{\ast}$ &$\Sigma_{b}B_{s}$
        &$\Lambda_{b}B_{s}^{\ast}$ &$\Lambda_{b}B_{s}$ \\ \hline
        1
            &${5/2}^{-}$    &11.524 &3.19, 0.24, -2.70 &1 & & & & & &1 & & & & & \\
            &               &11.378 & &$\ast$ & & & & & & & & & & & \\
            &${3/2}^{-}$    &11.654 &0.59, -0.14, -0.86 &0 &1 &0.0001 & & & &1.43 &1 &0.25 & & & \\
            &               &11.510 &2.60, 0.10, -2.41 &1.38 &1 &7.28 & & & &4.42 &1 &0.29 & & & \\
            &               &11.490 &1.96, 0.04, -1.88 &0.09 &1 &0.36 & & & &0.04 &1 &8.18 & & & \\
            &               &11.464 &2.51, 0.59, -1.33 &2.48 &1 &7.27 & & & &0.1 &1 &0.04 & & & \\
            &               &11.378 & &$\ast$ & & & & & & & & & & & \\
            &               &11.373 & & &$\ast$ & & & & & & & & & & \\
            &               &11.137 & & & &$\ast$ & & & & & & & & & \\
            &${1/2}^{-}$    &11.675 &0.12, -0.15, -0.42 &1 & &0.004 &0.004 & & &1 & &0.15 &0.03 & & \\
            &               &11.635 &0.27, 0.06, -0.15 &1 & &0.006 &0.004 & & &1 & &16.45 &8.56 & & \\
            &               &11.488 &1.45, 0.14, -1.18 &1 & &4.57 &3.06 & & &1 & &0.81 &0.2 & & \\
            &               &11.452 &-0.69, -0.39, -0.09 &1 & &47.08 &83.87 & & &1 & &1.45 &0.19 & & \\
            &               &11.445 &2.03, 0.59, -0.85 &1 & &5.93 &21.13 & & &0.05 & &1 &28.8 & & \\
            &               &11.376 & &$\ast$ & & & & & & & & & & & \\
            &               &11.137 & & & &$\ast$ & & & & & & & & & \\
            &               &11.131 & & & & &$\ast$ & & & & & & & & \\
        0
            &${5/2}^{-}$    &11.590 &0.24 &\multicolumn{6}{l|}{} \\
            &${3/2}^{-}$    &11.538 &0.24 & & & & &1 & & & & & &1 & \\
            &               &11.521 &0.40 & & & & &1 & & & & & &1 & \\
            &               &11.447 &0.19 & & & & &1 & & & & & &1 & \\
            &               &11.262 &-0.25 & & & & &1 & & & & & &1 & \\
            &               &11.094 & & & & & &$\ast$ & & & & & & & \\
            &${1/2}^{-}$    &11.505 &0.26 & & & & &1 &8.85 & & & & &1 &0.05 \\
            &               &11.439 &0.39 & & & & &1 &0.003 & & & & &1 &1.21 \\
            &               &11.429 &-0.24 & & & & &1 &27.57 & & & & &1 &1.29 \\
            &               &11.259 &-0.09 & & & & &1 &0.71 & & & & &1 &0.002 \\
            &               &11.210 &-0.08 & & & & &1 &3.24 & & & & &0.002 &1 \\
            &               &11.094 & & & & & &$\ast$ & & & & & & & \\
            &               &11.088 & & & & & & &$\ast$ & & & & & & \\
        \bottomrule[0.5pt]\bottomrule[1.5pt]
    \end{tabular}
\end{table*}

\renewcommand{\tabcolsep}{0.12cm}
\renewcommand{\arraystretch}{1.0}
\begin{table*}[!htbp]
    \caption{Calculated spectra (in GeV) of pentaquarks $nnsc\bar{c}$. 
    Magnetic moments are in unit of $\mu_N$, and organized in the order of $I_3=1, 0, -1$ for $I=1$, or $I_3=0$ for $I=0$.
    The bag radius $R_0$ is determined to be 5.78$\,$GeV$^{-1}$.
    The numbers below respective decay channels are ratios of partial width.
    The states denoted by asterisks couple strongly to scattering states.}
    \label{tab:nnscc}
    \begin{tabular}{cccc|cccccc|cccccc}
        \bottomrule[1.5pt]\bottomrule[0.5pt]
        \multirow{2}{*}{$I$} &\multirow{2}{*}{$J^{P}$}
        &\multicolumn{2}{l|}{$nnsc\bar{c}$} 
        &\multicolumn{6}{l|}{$nns\otimes c\bar{c}$} &\multicolumn{6}{l}{$nnc\otimes s\bar{c}$} \\
        & &$M$ &$\mu$ &$\Sigma^{\ast}J/\psi$ &$\Sigma^{\ast}\eta_{c}$
        &$\Sigma J/\psi$ &$\Sigma\eta_{c}$ &$\Lambda J/\psi$ &$\Lambda\eta_{c}$
        &$\Sigma_{c}^{\ast}D_{s}^{\ast}$ &$\Sigma_{c}^{\ast}D_{s}$ &$\Sigma_{c}D_{s}^{\ast}$ &$\Sigma_{c}D_{s}$
        &$\Lambda_{c}D_{s}^{\ast}$ &$\Lambda_{c}D_{s}$ \\ \hline
        1
            &${5/2}^{-}$    &4.767 &3.36, 0.27, -2.82 &1 & & & & & &1 & & & & & \\
            &               &4.690 & &$\ast$ & & & & & & & & & & & \\
            &${3/2}^{-}$    &4.863 &0.84, -0.08, -0.99 &0.0001 &1 &0.0004 & & & &1.18 &1 &0.15 & & & \\
            &               &4.751 &2.62, 0.42, -1.77 &1.16 &1 &2.03 & & & &10.27 &1 &0.07 & & & \\
            &               &4.691 & &$\ast$ & & & & & & & & & & & \\
            &               &4.683 &1.66, -0.46, -2.58 &7.26 &1 &0.3 & & & &0 &1 &28.38 & & & \\
            &               &4.651 &3.70, 0.88, -1.94 &0.003 &1 &0.006 & & & &0.0002 &1 &0.03 & & & \\
            &               &4.643 &2.99, 0.40, -2.19 &0.0005 &1 &0.001 & & & &0.02 &1 &0.73 & & & \\
            &               &4.451 & & & &$\ast$ & & & & & & & & & \\
            &${1/2}^{-}$    &4.920 &0.94, 0.51, 0.08 &1 & &0.002 &0.002 & & &1 & &0.08 &0.02 & & \\
            &               &4.814 &-0.29, -0.52, -0.74 &1 & &0.007 &0.002 & & &0.002 & &1 &0.36 & & \\
            &               &4.716 &1.41, 0.12, -1.16 &1 & &1.43 &0.26 & & &1 & &0.68 &0.03 & & \\
            &               &4.666 & &$\ast$ & & & & & & & & & & & \\
            &               &4.637 &-0.39, -0.01, 0.37 &1 & &0.17 &10.45 & & &1 & &6.32 &0.69 & & \\
            &               &4.594 &1.76, 0.16, -1.43 &0.0007 & &1 &2.14 & & &- & &1 &15.12 & & \\
            &               &4.445 & & & &$\ast$ & & & & & & & & & \\
            &               &4.402 & & & & &$\ast$ & & & & & & & & \\
        0
            &${5/2}^{-}$    &4.792 &0.27 &\multicolumn{6}{l|}{} \\
            &${3/2}^{-}$    &4.757 &0.59 & & & & &1 & & & & & &1 & \\
            &               &4.708 &0.19 & & & & &1 & & & & & &1 & \\
            &               &4.630 &0.12 & & & & &1 & & & & & &1 & \\
            &               &4.494 &-0.27 & & & & &1 & & & & & &1 & \\
            &               &4.406 & & & & & &$\ast$ & & & & & & & \\
            &${1/2}^{-}$    &4.703 &0.25 & & & & &1 &0.62 & & & & &1 &0.15 \\
            &               &4.619 &0.59 & & & & &1 &3.13 & & & & &1 &21.73 \\
            &               &4.592 &-0.42 & & & & &1 &3.57 & & & & &1 &0.02 \\
            &               &4.487 &-0.64 & & & & &1 &1.47 & & & & &1 &0.002 \\
            &               &4.402 & & & & & &$\ast$ & & & & & & & \\
            &               &4.358 & & & & & & &$\ast$ & & & & & & \\
            &               &4.353 &0.32 & & & & &1 &2.17 & & & & &- &1 \\
        \bottomrule[0.5pt]\bottomrule[1.5pt]
    \end{tabular}
\end{table*}

In the subsequent analysis, we explore the considerable hidden-heavy pentaquarks $nnsQ\bar{Q}$ with a single strange flavor.
For these systems, the color-spin bases expand to eight for $I=1$ and seven for $I=0$, detailed in Appendix A.
Consequently, numerical calculations must rely on the perturbative method of MIT bag model, where we variate Eq.(\ref{equ:mass}) 
by treating $M_{CMI}$ matrix as perturbation. To investigate the decay channels, we employ the relations
\begin{equation}
    \begin{aligned}
        &\gamma_{\Sigma^{\ast}\Upsilon} = \gamma_{\Sigma^{\ast}\eta_b} = \gamma_{\Sigma\Upsilon} = \gamma_{\Sigma\eta_b}, \quad
        \gamma_{\Lambda\Upsilon} = \gamma_{\Lambda\eta_b}, \\
        &\gamma_{\Sigma^{\ast}J/\psi} = \gamma_{\Sigma^{\ast}\eta_c} = \gamma_{\Sigma J/\psi} = \gamma_{\Sigma\eta_c}, \quad
        \gamma_{\Lambda J/\psi} = \gamma_{\Lambda\eta_c}, \\
        &\gamma_{\Sigma_b^{\ast}B_s^{\ast}} = \gamma_{\Sigma_b^{\ast}B_s} = \gamma_{\Sigma_b B_s^{\ast}} = \gamma_{\Sigma_b B_s}, \quad
        \gamma_{\Lambda_b B_s^{\ast}} = \gamma_{\Lambda_b B_s}, \\
        &\gamma_{\Sigma_c^{\ast}D_s^{\ast}} = \gamma_{\Sigma_c^{\ast}D_s} = \gamma_{\Sigma_c D_s^{\ast}} = \gamma_{\Sigma_c D_s}, \quad
        \gamma_{\Lambda_c D_s^{\ast}} = \gamma_{\Lambda_c D_s},
    \end{aligned}
\end{equation}
applied to Eq.(\ref{equ:partialwidth}).
The results of mass spectrum, magnetic moment and ratios of partial widths are presented in Tables \ref{tab:nnsbb} and \ref{tab:nnscc}.

The scattering states entering for each spin-parity should be excluded first. Here, we denote a pentaquark state as $P_{Qs}(I,J^P,M)$ in this section.
For subsystems with $I=0$, we identify several scattering states like $P_{bs}(0,3/2^-,11.094)$, $P_{bs}(0,1/2^-,11.094)$ and $P_{bs}(0,1/2^-,11.088)$ for bottom sector,
and $P_{cs}(0,3/2^-,4.406)$, $P_{cs}(0,1/2^-,4.402)$ and $P_{cs}(0,1/2^-,4.358)$ for charm sector.
In the case of $I=1$, thirteen more scattering states are explicitly marked with asterisks.
It's noteworthy that the lightest $P_{cs}$ pentaquark $P_{cs}(0,1/2^-,4.353)$ remains as a tightly bound state.
Accordingly, we have $P_{bs}$ pentaquarks ranging from 11.210$\,$GeV to 11.675$\,$GeV and $P_{cs}$ from 4.353$\,$GeV to 4.920$\,$GeV,
each with significant mass splittings of 465$\,$MeV and 567$\,$MeV, respectively, due to the presence of a single strange flavor.

Regarding the $P_{cs}$ pentaquark candidates, LHCb Collaboration has reported two resonances in $J/\psi\Lambda$ final states:
\begin{equation}
    \begin{aligned}
        &P_{\psi s}^{\Lambda}(4338)^0 \ M=4338.2\,\textrm{MeV} \ \Gamma=7.0\,\textrm{MeV} \text{\cite{LHCb:2022ogu}}, \\
        &P_{cs}(4459)^0 \ M=4458.8\,\textrm{MeV} \ \Gamma=17.3\,\textrm{MeV} \text{\cite{LHCb:2020jpq}},
    \end{aligned}
\end{equation}
with preferred spin-parity $1/2^-$ of the former.
By examining the corresponding threshold of $P_{\psi s}^{\Lambda}(4338)^0$, we find that its mass of 4338.2$\,$MeV
is slightly higher than $\Xi_c D$, below $\Xi_c^{\prime}D$ by 108$\,$MeV and below $\Xi_c D^{\ast}$ by 140$\,$MeV.
Despite the deep binding of over 100$\,$MeV, suggesting it may not be a molecular state,
$P_{\psi s}^{\Lambda}(4338)^0$ can be described as a compact pentaquark $P_{cs}(0,1/2^-,4.353)$ with a magnetic moment of $0.32\mu_N$
and ratio of partial widths
\begin{equation}
    \frac{\Gamma(P_{cs}(0,1/2^-,4.353) \to \Lambda\eta_c)}{\Gamma(P_{cs}(0,1/2^-,4.353) \to \Lambda J/\psi)} = 2.17.
\end{equation}
The spin-parity of $P_{cs}(4459)^0$ has not been determined and could be either $3/2^-$ or $1/2^-$, corresponding to predicted
$P_{cs}(0,3/2^-,4.494)$ and $P_{cs}(0,1/2^-,4.487)$, respectively. We are looking forward to further experiments on 
quantum numbers and decay channel $\Lambda\eta_c$, for confirming the two-peak hypothesis of $P_{cs}(4459)^0$ 
dissociating into 4454.9$\,$MeV and 4467.8$\,$MeV \cite{LHCb:2020jpq}.
Additionally, the Fig 3(b) of Ref.\cite{LHCb:2020jpq} suggests the existence of several $P_{cs}$ states in the 4.6-4.9$\,$GeV region,
as predicted in Table \ref{tab:nnscc}.

In this work, our focus extends to the examination of $P_{bs}$ pentaquarks, considering their magnetic moments and decay channels.
Due to the lack of evidence and production mechanism, there is no comparison of them to experimental reports.
Nonetheless, similar to the present of a single massive strange flavor in $P_{cs}$ counterparts, 
the bottom flavor can exhibit deep binding and suppression of relativistic effects compared to strange and charm flavors.
Theoretical predictions, especially those related to magnetic moments and decay channels, 
can serve as guidance for expected findings of compact $P_{bs}$ pentaquarks.

\subsection{The $ssnQ\bar{Q}$ systems}
\label{sec:ssnQQ}

\renewcommand{\tabcolsep}{0.15cm}
\renewcommand{\arraystretch}{1.0}
\begin{table*}[!htbp]
    \caption{Calculated spectra (in GeV) of pentaquarks $ssnb\bar{b}$. 
    Magnetic moments are in unit of $\mu_N$, and organized in the order of $I_3=1/2, -1/2$ for $I=1/2$.
    The bag radius $R_0$ is determined to be 5.53$\,$GeV$^{-1}$.
    The numbers below respective decay channels are ratios of partial width.
    The states denoted by asterisks couple strongly to scattering states.}
    \label{tab:ssnbb}
    \begin{tabular}{ccc|cccc|cccc|cccc}
        \bottomrule[1.5pt]\bottomrule[0.5pt]
        \multirow{2}{*}{$J^{P}$}
        &\multicolumn{2}{l|}{$ssnb\bar{b}$} 
        &\multicolumn{4}{l|}{$ssn\otimes b\bar{b}$} &\multicolumn{4}{l|}{$nsb\otimes s\bar{b}$}
        &\multicolumn{4}{l}{$ssb\otimes n\bar{b}$} \\
        &$M$ &$\mu$ &$\Xi^{\ast}\Upsilon$ &$\Xi^{\ast}\eta_{b}$ &$\Xi\Upsilon$ &$\Xi\eta_{b}$
        &$\Xi_{b}^{\ast}B_{s}^{\ast}$ &$\Xi_{b}^{\ast}B_{s}$ &$\Xi_{b}B_{s}^{\ast}$ &$\Xi_{b}B_{s}$
        &$\Omega_{b}^{\ast}B^{\ast}$ &$\Omega_{b}^{\ast}B$ &$\Omega_{b}B^{\ast}$ &$\Omega_{b}B$ \\ \hline
            ${5/2}^{-}$     &11.669 &0.49, -2.46 &1 & & & &1 & & & &1 & & & \\
                            &11.524 & &$\ast$ & & & & & & & & & & & \\
            ${3/2}^{-}$     &11.747 &-0.26, -0.79 &0 &1 &0.0001 & &1.54 &1 &0.3 & &5.4 &2.43 &1 & \\
                            &11.651 &0.51, -2.43 &7.54 &1 &43.94 & &3.6 &1 &1.79 & &1.46 &0.67 &1 & \\
                            &11.636 &0.61, -1.80 &0.0007 &1 &0.003 & &0.28 &1 &2.79 & &0.05 &0.52 &1 & \\
                            &11.561 &0.32, -0.90 &0.21 &1 &0.19 & &0.54 &1 &0.16 & &2.73 &8.89 &1 & \\
                            &11.524 & &$\ast$ & & & & & & & & & & & \\
                            &11.518 & & &$\ast$ & & & & & & & & & & \\
                            &11.283 & & & &$\ast$ & & & & & & & & & \\
            ${1/2}^{-}$     &11.766 &-0.22, -0.40 &1 & &0.003 &0.003 &1 & &0.01 &0.12 &40.84 & &5.93 &1 \\
                            &11.729 &0.06, -0.13 &1 & &0.005 &0.002 &1 & &0.04 &0.003 &0.21 & &2.91 &1 \\
                            &11.620 &0.41, -1.32 &1 & &1.12 &13.65 &1 & &0.34 &0.02 &0.33 & &0.51 &1 \\
                            &11.555 &0.61, -0.81 &1 & &2.53 &0.03 &0.04 & &1 &1.71 &0.98 & &0.16 &1 \\
                            &11.545 &-0.37, 0.19 &1 & &0.39 &4.37 &0.39 & &1 &0.67 &0.14 & &0.96 &1 \\
                            &11.521 & &$\ast$ & & & & & & & & & & & \\
                            &11.283 & & & &$\ast$ & & & & & & & & & \\
                            &11.277 & & & & &$\ast$ & & & & & & & & \\
        \bottomrule[0.5pt]\bottomrule[1.5pt]
    \end{tabular}
\end{table*}

\renewcommand{\tabcolsep}{0.15cm}
\renewcommand{\arraystretch}{1.0}
\begin{table*}[!htbp]
    \caption{Calculated spectra (in GeV) of pentaquarks $ssnc\bar{c}$. 
    Magnetic moments are in unit of $\mu_N$, and organized in the order of $I_3=1/2, -1/2$ for $I=1/2$.
    The bag radius $R_0$ is determined to be 5.79$\,$GeV$^{-1}$.
    The numbers below respective decay channels are ratios of partial width.
    The states denoted by asterisks couple strongly to scattering states.}
    \label{tab:ssncc}
    \begin{tabular}{ccc|cccc|cccc|cccc}
        \bottomrule[1.5pt]\bottomrule[0.5pt]
        \multirow{2}{*}{$J^{P}$}
        &\multicolumn{2}{l|}{$ssnc\bar{c}$} 
        &\multicolumn{4}{l|}{$ssn\otimes c\bar{c}$} &\multicolumn{4}{l|}{$nsc\otimes s\bar{c}$}
        &\multicolumn{4}{l}{$ssc\otimes n\bar{c}$} \\
        &$M$ &$\mu$ &$\Xi^{\ast}J/\psi$ &$\Xi^{\ast}\eta_{c}$ &$\Xi J/\psi$ &$\Xi\eta_{c}$
        &$\Xi_{c}^{\ast}D_{s}^{\ast}$ &$\Xi_{c}^{\ast}D_{s}$ &$\Xi_{c}D_{s}^{\ast}$ &$\Xi_{c}D_{s}$
        &$\Omega_{c}^{\ast}D^{\ast}$ &$\Omega_{c}^{\ast}D$ &$\Omega_{c}D^{\ast}$ &$\Omega_{c}D$ \\ \hline
            ${5/2}^{-}$     &4.914 &0.54, -2.56 &1 & & & &1 & & & &1 & & & \\
                            &4.836 & &$\ast$ & & & & & & & & & & & \\
            ${3/2}^{-}$     &4.968 &-0.11, -0.93 &0.0001 &1 &0.0007 & &1.49 &1 &0.19 & &9.01 &2.65 &1 & \\
                            &4.881 &0.90, -2.04 &2.74 &1 &3.52 & &6.67 &1 &0.47 & &3.8 &1.34 &1 & \\
                            &4.837 &0.48, -2.13 &30.3 &1 &0.05 & &1.18 &1 &4.53 & &0.02 &0.22 &1 & \\
                            &4.834 &0.64, -2.07 &35.37 &1 &0.14 & &0.07 &1 &4.12 & &0.01 &0.0005 &1 & \\
                            &4.792 & & &$\ast$ & & & & & & & & & & \\
                            &4.760 &-0.28, -0.95 &0.05 &1 &0.62 & &0.01 &1 &0.0001 & &- &1 &0 & \\
                            &4.597 & & & &$\ast$ & & & & & & & & & \\
            ${1/2}^{-}$     &5.021 &0.44, 0.08 &1 & &0.002 &0.001 &1 & &0.02 &0.12 &67.1 & &4.23 &1 \\
                            &4.925 &-0.37, -0.71 &1 & &0.004 &0.0004 &1 & &0.06 &0.005 &0.23 & &8.31 &1 \\
                            &4.831 &0.39, -1.14 &1 & &0.28 &0.20 &1 & &2.91 &0.48 &5.28 & &2.65 &1 \\
                            &4.808 &0.31, -1.33 &1 & &0.002 &0.003 &1 & &2.27 &3.07 &0.01 & &0.004 &1 \\
                            &4.754 &0.68, 0.30 &1 & &7.09 &20.99 &- & &1 &18.33 &- & &3.27 &1 \\
                            &4.727 &-0.68, -1.16 &0.0006 & &1 &2.95 &- & &1 &0.004 &- & &0.008 &1 \\
                            &4.592 & & & &$\ast$ & & & & & & & & & \\
                            &4.548 & & & & &$\ast$ & & & & & & & & \\
        \bottomrule[0.5pt]\bottomrule[1.5pt]
    \end{tabular}
\end{table*}

In analogy to the isovector $nnsQ\bar{Q}$ system, a corresponding system can be constructed by reversing the flavors of $n$ and $s$,
resulting in the $ssnQ\bar{Q}$ system with seven and eight color-spin bases for $J^P=3/2^-$ and $1/2^-$, respectively.
The calculations are performed using the perturbative method of MIT bag model, 
along with the application of certain relations for partial width study, as expressed below:
\begin{equation}
    \begin{aligned}
        &\gamma_{\Xi^{\ast}\Upsilon} = \gamma_{\Xi^{\ast}\eta_b} = \gamma_{\Xi\Upsilon} = \gamma_{\Xi\eta_b}, 
        \gamma_{\Xi^{\ast}J/\psi} = \gamma_{\Xi^{\ast}\eta_c} = \gamma_{\Xi J/\psi} = \gamma_{\Xi\eta_c}, \\
        &\gamma_{\Xi_b^{\ast}B_s^{\ast}} = \gamma_{\Xi_b^{\ast}B_s} = \gamma_{\Xi_b B_s^{\ast}} = \gamma_{\Xi_b B_s}, 
        \gamma_{\Xi_c^{\ast}D_s^{\ast}} = \gamma_{\Xi_c^{\ast}D_s} = \gamma_{\Xi_c D_s^{\ast}} = \gamma_{\Xi_c D_s}, \\
        &\gamma_{\Omega_b^{\ast}B^{\ast}} = \gamma_{\Omega_b^{\ast}B} = \gamma_{\Omega_b B^{\ast}} = \gamma_{\Omega_b B}, 
        \gamma_{\Omega_c^{\ast}D^{\ast}} = \gamma_{\Omega_c^{\ast}D} = \gamma_{\Omega_c D^{\ast}} = \gamma_{\Omega_c D}. \\
    \end{aligned}
\end{equation}
The corresponding results are presented in Tables \ref{tab:ssnbb} and \ref{tab:ssncc}.

In this section, we will designate the pentaquark $ssnQ\bar{Q}$ by the symbol $P_{Qss}(J^P,M)$.
Beyond the scattering states, our predictions encompass ten $P_{bss}$ states ranging from 11.545$\,$GeV to 11.747$\,$GeV and twelve $P_{css}$ states
spanning 4.727$\,$GeV to 4.968$\,$GeV. These states exhibit mass differences of 202$\,$MeV and 241$\,$MeV, respectively.
Through the $ssn\otimes Q\bar{Q}$ coupling, there are several states can be found dominantly in channel $\Xi^{\ast} J/\psi$ ($\Xi^{\ast}\Upsilon$), 
including $P_{css}(1/2^-,4.808)$, $P_{css}(1/2^-,4.925)$ and $P_{css}(1/2^-,5.021)$ for $c$-sector, and $P_{bss}(1/2^-,11.729)$, $P_{bss}(1/2^-,11.766)$
for $b$-sector. The lightest $P_{css}$ pentaquark around 4.7$\,$GeV is expected to be firstly reported in experiments
either in system $\Xi J/\psi$, $\Xi\eta_c$, $\Xi_c D_s^{\ast}$ or $\Omega_c D^{\ast}$.

The formation of a molecular state for a $P_{Qss}$ pentaquark proves challenging due to the exchange of mesons
$s\bar{s}$, $c\bar{c}$ and $b\bar{b}$, which only offer short-range interactions among constituent hadrons.
Additionally, the superthreshold phenomenon further violates this scenario by positive potentials, implying repulsive force between hadrons.
Consequently, we infer that the hidden-heavy pentaquarks with strangeness might be discovered as compact states,
arisen from the unsatisfied requirements of the molecular scenario.

\subsection{The $sssQ\bar{Q}$ systems}
\label{sec:sssQQ}

\renewcommand{\tabcolsep}{0.5cm}
\renewcommand{\arraystretch}{1.0}
\begin{table*}[!htbp]
    \caption{Calculated spectra (in GeV) of pentaquarks $sssb\bar{b}$. Bag radius $R_0$ is in GeV$^{-1}$. Magnetic moments are in unit of $\mu_N$.
    The numbers below respective decay channels are ratios of partial width.
    The states denoted by asterisks couple strongly to scattering states.}
    \label{tab:sssbb}
    \begin{tabular}{cccc|cc|cccc}
        \bottomrule[1.5pt]\bottomrule[0.5pt]
        \multirow{2}{*}{$J^{P}$} 
        &\multicolumn{3}{l|}{$sssb\bar{b}$} &\multicolumn{2}{l|}{$sss\otimes b\bar{b}$}
        &\multicolumn{4}{l}{$ssb\otimes s\bar{b}$} \\
        &$R_{0}$ &$M$ &$\mu$ 
        &$\Omega\Upsilon$ &$\Omega\eta_{b}$ &$\Omega_{b}^{\ast}B_{s}^{\ast}$
        &$\Omega_{b}^{\ast}B_{s}$ &$\Omega_{b}B_{s}^{\ast}$ &$\Omega_{b}B_{s}$ \\ \hline
            ${5/2}^{-}$    &5.63   &11.673 & &$\ast$ & & & & & \\
            ${3/2}^{-}$    &5.65   &11.841 &-0.76 &0 &1 &1 &0.56 &0.18 & \\
                           &5.63   &11.673 & &$\ast$ & & & & & \\
                           &5.62   &11.668 & & &$\ast$ & & & & \\
            ${1/2}^{-}$    &5.69   &11.860 &-0.39 &1 & &1 & &0.14 &0.03 \\
                           &5.62   &11.823 &-0.11 &1 & &1 & &16.21 &6.56 \\
                           &5.62   &11.671 & &$\ast$ & & & & & \\
        \bottomrule[0.5pt]\bottomrule[1.5pt]
    \end{tabular}
\end{table*}

\renewcommand{\tabcolsep}{0.5cm}
\renewcommand{\arraystretch}{1.0}
\begin{table*}[!htbp]
    \caption{Calculated spectra (in GeV) of pentaquarks $sssc\bar{c}$. Bag radius $R_0$ is in GeV$^{-1}$. Magnetic moments are in unit of $\mu_N$.
    The numbers below respective decay channels are ratios of partial width.
    The states denoted by asterisks couple strongly to scattering states.}
    \label{tab:ssscc}
    \begin{tabular}{cccc|cc|cccc}
        \bottomrule[1.5pt]\bottomrule[0.5pt]
        \multirow{2}{*}{$J^{P}$} 
        &\multicolumn{3}{l|}{$sssc\bar{c}$} &\multicolumn{2}{l|}{$sss\otimes c\bar{c}$} 
        &\multicolumn{4}{l}{$ssc\otimes s\bar{c}$} \\
        &$R_{0}$ &$M$ &$\mu$ 
        &$\Omega J/\psi$ &$\Omega\eta_{c}$ &$\Omega_{c}^{\ast}D_{s}^{\ast}$
        &$\Omega_{c}^{\ast}D_{s}$ &$\Omega_{c}D_{s}^{\ast}$ &$\Omega_{c}D_{s}$ \\ \hline
            ${5/2}^{-}$    &5.90   &4.987 & &$\ast$ & & & & & \\
            ${3/2}^{-}$    &5.93   &5.076 &-0.91 &0 &1 &1 &0.45 &0.09 & \\
                           &5.90   &4.987 & &$\ast$ & & & & & \\
                           &5.80   &4.940 & & &$\ast$ & & & & \\
            ${1/2}^{-}$    &6.00   &5.126 &0.05  &1 & &1 & &0.05 &0.01 \\
                           &5.88   &5.037 &-0.64 &1 & &1 & &326.17 &33.47 \\
                           &5.82   &4.953 &-1.20 &1 & &1 & &0.87 &31.37 \\
        \bottomrule[0.5pt]\bottomrule[1.5pt]
    \end{tabular}
\end{table*}

Finally, when three strange quarks are present, the light degrees of freedom in $sssQ\bar{Q}$ system are all identical in flavor. 
The strong symmetry of the wavefunction constrains the system to three eigenstates with spin-parity $3/2^-$ or $1/2^-$,
and due to the exclusion of scattering states, our predictions for compact pentaquarks are more limited.
Similar to the previous cases, the color-spin bases outlined in Appendix A, along with the relations
\begin{equation}
    \begin{aligned}
        &\gamma_{\Omega\Upsilon} = \gamma_{\Omega\eta_b},\quad \gamma_{\Omega J/\psi} = \gamma_{\Omega\eta_c}, \\
        &\gamma_{\Omega_b^{\ast}B_s^{\ast}} = \gamma_{\Omega_b^{\ast}B_s} = \gamma_{\Omega_b B_s^{\ast}} = \gamma_{\Omega_b B_s}, \\
        &\gamma_{\Omega_c^{\ast}D_s^{\ast}} = \gamma_{\Omega_c^{\ast}D_s} = \gamma_{\Omega_c D_s^{\ast}} = \gamma_{\Omega_c D_s},
    \end{aligned}
\end{equation}
are employed to evaluate hadron properties of $sssQ\bar{Q}$. The numerical results are detailed in Tables \ref{tab:sssbb} and \ref{tab:ssscc}. 

In this context, we characterize a $sssQ\bar{Q}$ pentaquark state using the notation $P_{Qsss}(J^P,M)$.
Notably, the bottom system features $P_{bsss}(1/2^-,11.823)$, $P_{bsss}(1/2^-,11.860)$ and $P_{bsss}(3/2^-,11.841)$.
The lightest and heaviest states of $P_{csss}$ are $P_{csss}(1/2^-,4.953)$ and $P_{csss}(1/2^-,5.126)$, respectively.
We predict that, the $P_{Qsss}$ states with $J^P=1/2^-$ can be discovered in channel $\Omega\Upsilon$ ($\Omega J/\psi$), 
and that of $3/2^-$ are anticipated in $\Omega\eta_b$ ($\Omega\eta_c$). 
Additionally, we explore the ratios of partial widths in the coupling $ssQ\otimes s\bar{Q}$,
where the state $P_{csss}(1/2^-,5.037)$ exhibits a dominant channel in $\Omega_c D_s^{\ast}$.
These features provide insights for predicting the spin-parity of pentaquarks based on reported decay channels.

\section{Summary}
\label{sec:summary}
In this study, we conducted a systematic investigation of hidden-heavy pentaquarks with strangeness $S=0,-1,-2,-3$, employing the unified framework of MIT bag model. 
Inspired by similar analysis applied to light-flavor baryons, a heavy $Q\bar{Q}$ in pentaquark configurations introduces a compact scenario characterized by deep binding.
With the help of color-spin bases expressed in terms of Young tableau and Young-Yamanouchi bases for $SU_c(3)\otimes SU_s(2)$ group, 
we computed the masses and magnetic moments of possible pentaquark states. In the baryon-meson coupling, the color-spin wavefunctions
are dissociated into color-singlet and color-octet components, providing a foundation for exploring the stability and ratios of partial widths.

For the non-strange $P_c$ and $P_b$ states with mass ranges of 4.45-4.82$\,$GeV and 11.31-11.58$\,$GeV respectively, 
the reported resonances fall below these ranges, suggesting a likelihood of molecular states. However, we anticipate that compact $P_c$
states with a spin-parity $1/2^-$ can be found above 4.45$\,$GeV in both $NJ/\psi$ and $N\eta_c$ systems. The finding of strange
$P_{cs}$ candidate $P_{\psi s}^{\Lambda}(4338)^0$ aligns with our predictions regarding the mass spectrum and spin-parity. Meanwhile, the $P_{cs}(4459)^0$ is expected 
to contain two substructures around 4459$\,$MeV with $J^P=3/2^-$ and $1/2^-$, as indicated by both experimental fits and our computations.
We eagerly await further reports on decay channels $N\eta_c$ and $\Lambda\eta_c$, for which we have studied the ratios of partial widths in this work.

Simultaneously, pentaquarks with strangeness $S=-2$ and $-3$ as well as those in the bottom sector are regarded to be compact.
There are two reasons: (1). Molecular scenario requires long-range mesons exchange, for this one must establish
interactions between constituent hadrons involving light non-strange mesons like $\pi$, $\omega$, $\rho$. 
However, in the $ssnQ\bar{Q}$ and $sssQ\bar{Q}$ systems, such quark constituents are absent.
Additionally, the superthreshold states conflict with the attractive potential inferred from negative binding.
(2). The inclusion of massive flavors, such as strange, charm, and bottom, can exhibit deep binding between the corresponding quarks, leading to substantial suppressions of relativistic effects.
Therefore, the searching for compact $P_{Qss}$ and $P_{Qsss}$ states, especially $P_{css}$ around 4.7$\,$GeV, will be reasonable.

In addition to the mass spectrum and partial widths, the magnetic moments serve as crucial factor in the prediction of hadronic states.
Our researches involved the calculation of the magnetic moment for each pentaquark, arranged by the third component of isospin, which corresponds to the charge of the final states.
We anticipate that our predictions of hadron properties and decay behaviors will provide valuable guidance in searching for hidden-heavy pentaquarks.

\medskip \textbf{ACKNOWLEDGMENTS}

W. Z. thanks Hong-Tao An for useful discussions on wavefunctions and partial width, 
and Wen-Nian Liu for valuable comments about pentaquark candidates. 
D. J. is supported by the National Natural Science Foundation of China under Grant No. 12165017.

\medskip
\section*{Appendix A: Color and Spin Wavefunctions}
\label{apd:basis}
\setcounter{equation}{0}
\renewcommand{\theequation}{A\arabic{equation}}

For numerical calculations of masses and magnetic moments, it's essential to employ color-spin wavefunctions to describe the chromomagnetic structure 
of a given pentaquark state. In the context of hidden-heavy pentaquarks with three light degrees of freedom, indicating flavor symmetry up to 
three identical quarks, the Young tableau is employed to represent color and spin wavefunctions in terms of the Young-Yamanouchi bases.
The use of Young tableau and the SU(3) permutation group for color representations and its application to various pentaquark systems has been 
discussed and implemented in prior works. For detailed discussions, see Refs.\cite{Zhang:2023hmg,An:2020vku,An:2020jix,An:2021vwi,An:2022fvs}.

The color wavefunctions in the configuration $q_1q_2q_3q_4\bar{q}_5$
constrained by overall color confinement, are selected in color-singlet through the application of the Young tableau. 
These wavefunctions are expressed as follows
\renewcommand{\tabcolsep}{0.1cm}
\renewcommand{\arraystretch}{1}
\begin{equation}
    {\begin{tabular}{|c|c|}
        \hline
        1 & 2 \\
        \cline{1-2}
        \multicolumn{1}{|c|}{3} \\
        \cline{1-1}
        \multicolumn{1}{|c|}{4} \\
        \cline{1-1}
    \end{tabular}
    \otimes\bar{5}}_{\begin{tabular}{|c|} \multicolumn{1}{c}{$\phi_{1}$}\end{tabular}},
    {\begin{tabular}{|c|c|}
        \hline
        1 & 3 \\
        \cline{1-2}
        \multicolumn{1}{|c|}{2} \\
        \cline{1-1}
        \multicolumn{1}{|c|}{4}  \\
        \cline{1-1}
    \end{tabular}
    \otimes\bar{5}}_{\begin{tabular}{|c|} \multicolumn{1}{c}{$\phi_{2}$}\end{tabular}},
    {\begin{tabular}{|c|c|}
        \hline
        1 & 4 \\
        \cline{1-2}
        \multicolumn{1}{|c|}{2} \\
        \cline{1-1}
        \multicolumn{1}{|c|}{3} \\
        \cline{1-1}
    \end{tabular}
    \otimes\bar{5}}_{\begin{tabular}{|c|} \multicolumn{1}{c}{$\phi_{3}$}\end{tabular}}.
    \label{equ:colorbases}
\end{equation}
It's notable that, in this study, particular attention is given to the third color basis denoted as $\phi_3$,
which serves to couple the color-singlet wavefunctions of the baryon and meson, resulting in $(q_1q_2q_3)\otimes(q_4\bar{q}_5)$.
This specific coupling corresponds to the unstable scattering state and final states of the OZI-superallowed decay mode.

Similarly, spin wavefunctions can be expressed in terms of Young tableau [5],[4,1], and [3,2] and classified into spin multiplets.
For the pentaquark with spin $J=5/2$, there is one basis
\renewcommand{\tabcolsep}{0.1cm}
\renewcommand{\arraystretch}{1}
\begin{equation}
    \begin{tabular}{|c|c|c|c|c|} 
        \hline
        1 & 2 & 3 & 4 & 5 \\ 
        \hline
    \end{tabular}_{\begin{tabular}{c}$\chi _{1}$\end{tabular}}.
    \label{equ:spinbases1}
\end{equation}
In the case of the $J=3/2$, the spin wavefunctions are
\renewcommand{\tabcolsep}{0.1cm}
\renewcommand{\arraystretch}{1}
\begin{equation}
    \begin{aligned}
        \begin{tabular}{|c|c|c|c|}
            \hline 
            1 & 2 & 3 & 4 \\
            \cline{1-4}
            \multicolumn{1}{|c|}{5} \\
            \cline{1-1}
            \end{tabular}_{\begin{tabular}{|c|}\multicolumn{1}{c}{$\chi_{2}$}\end{tabular}},
        \begin{tabular}{|c|c|c|c|}
            \hline 
            1 & 2 & 3 & 5 \\
            \cline{1-4}
            \multicolumn{1}{|c|}{4} \\ 
            \cline{1-1}
        \end{tabular}_{\begin{tabular}{|c|}\multicolumn{1}{c}{$\chi_{3}$}\end{tabular}}, \\
        \begin{tabular}{|c|c|c|c|}
            \hline
            1 & 2 & 4 & 5 \\
            \cline{1-4}
            \multicolumn{1}{|c|}{3} \\ 
            \cline{1-1}
        \end{tabular}_{\begin{tabular}{|c|}\multicolumn{1}{c}{$\chi_{4}$}\end{tabular}}, 
        \begin{tabular}{|c|c|c|c|}
            \hline 
            1 & 3 & 4 & 5 \\ 
            \cline{1-4} 
            \multicolumn{1}{|c|}{2} \\ 
            \cline{1-1}
        \end{tabular}_{\begin{tabular}{|c|}\multicolumn{1}{c}{$\chi_{5}$}\end{tabular}}, 
    \end{aligned}
    \label{equ:spinbases2}
\end{equation}
and for $J=1/2$, they become
\renewcommand{\tabcolsep}{0.1cm}
\renewcommand{\arraystretch}{1}
\begin{equation}
    \begin{aligned}
        \begin{tabular}{|c|c|c|} 
            \hline 
            1 & 2 & 3 \\ 
            \cline{1-3}
            4 & 5 \\ 
            \cline{1-2} 
        \end{tabular}_{\begin{tabular}{|c|}\multicolumn{1}{c}{$\chi_{6}$}\end{tabular}}, 
        \begin{tabular}{|c|c|c|}
            \hline 
            1 & 2 & 4 \\ 
            \cline{1-3} 
            3 & 5 \\ 
            \cline{1-2}
        \end{tabular}_{\begin{tabular}{|c|}\multicolumn{1}{c}{$\chi_{7}$}\end{tabular}}, 
        \begin{tabular}{|c|c|c|}
            \hline 
            1 & 3 & 4 \\ 
            \cline{1-3} 
            2 & 5 \\ 
            \cline{1-2}
        \end{tabular}_{\begin{tabular}{|c|}\multicolumn{1}{c}{$\chi_{8}$}\end{tabular}}, \\
        \begin{tabular}{|c|c|c|}
            \hline 
            1 & 2 & 5 \\ 
            \cline{1-3} 
            3 & 4 \\
            \cline{1-2} 
        \end{tabular}_{\begin{tabular}{|c|}\multicolumn{1}{c}{$\chi_{9}$}\end{tabular}}, 
        \begin{tabular}{|c|c|c|}
            \hline 
            1 & 3 & 5 \\ 
            \cline{1-3} 
            2 & 4 \\ 
            \cline{1-2}
        \end{tabular}_{\begin{tabular}{|c|}\multicolumn{1}{c}{$\chi_{10}$}\end{tabular}}. 
    \end{aligned}
    \label{equ:spinbases3}
\end{equation}
In addition to their role in calculating chromomagnetic interactions, these spin wavefunctions are sufficient to derive matrix elements of magnetic moments
as shown in Appendix B, with the help of the following expanding bases,
\begin{align}
    \chi_{1} &= \uparrow\uparrow\uparrow\uparrow\uparrow, \nonumber\\[3mm]
    \chi_{2} &= \frac{2}{\sqrt{5}}\uparrow\uparrow\uparrow\uparrow\downarrow
        -\frac{\sqrt{5}}{10}\left(\uparrow\uparrow\uparrow\downarrow\uparrow + \uparrow\uparrow\downarrow\uparrow\uparrow
        +\uparrow\downarrow\uparrow\uparrow\uparrow + \downarrow\uparrow\uparrow\uparrow\uparrow\right), \nonumber\\ 
    \chi_{3} &= \frac{\sqrt{3}}{2}\uparrow\uparrow\uparrow\downarrow\uparrow - \frac{1}{2\sqrt{3}}\left(\uparrow\uparrow\downarrow\uparrow\uparrow
        +\uparrow\downarrow\uparrow\uparrow\uparrow + \downarrow\uparrow\uparrow\uparrow\uparrow\right), \nonumber\\ 
    \chi_{4} &= \frac{1}{\sqrt{6}}\left(2\uparrow\uparrow\downarrow\uparrow\uparrow - \uparrow\downarrow\uparrow\uparrow\uparrow 
        -\downarrow\uparrow\uparrow\uparrow\uparrow\right), \nonumber\\ 
    \chi_{5} &= \frac{1}{\sqrt{2}}\left(\uparrow\downarrow\uparrow\uparrow\uparrow - \downarrow\uparrow\uparrow\uparrow\uparrow\right), \nonumber\\ 
    \chi_{6} &= \frac{1}{3\sqrt{2}}\left(\uparrow\downarrow\downarrow\uparrow\uparrow + \downarrow\uparrow\downarrow\uparrow\uparrow
        +\downarrow\downarrow\uparrow\uparrow\uparrow - \uparrow\uparrow\downarrow\downarrow\uparrow
        -\uparrow\downarrow\uparrow\downarrow\uparrow \right. \nonumber\\
        &\left. -\downarrow\uparrow\uparrow\downarrow\uparrow
        -\uparrow\uparrow\downarrow\uparrow\downarrow - \uparrow\downarrow\uparrow\uparrow\downarrow
        -\downarrow\uparrow\uparrow\uparrow\downarrow\right)
        +\frac{1}{\sqrt{2}}\uparrow\uparrow\uparrow\downarrow\downarrow, \nonumber\\ 
    \chi_{7} &= \frac{1}{3}\left(2\uparrow\uparrow\downarrow\uparrow\downarrow - \uparrow\downarrow\uparrow\uparrow\downarrow
        -\downarrow\uparrow\uparrow\uparrow\downarrow - \uparrow\uparrow\downarrow\downarrow\uparrow + \downarrow\downarrow\uparrow\uparrow\uparrow\right) \nonumber\\
        &+\frac{1}{6}\left(\uparrow\downarrow\uparrow\downarrow\uparrow - \uparrow\downarrow\downarrow\uparrow\uparrow
        -\downarrow\uparrow\downarrow\uparrow\uparrow + \downarrow\uparrow\uparrow\downarrow\uparrow\right), \nonumber\\ 
    \chi_{8} &= \frac{1}{\sqrt{3}}\left(\uparrow\downarrow\uparrow\uparrow\downarrow - \downarrow\uparrow\uparrow\uparrow\downarrow\right)
        -\frac{1}{2\sqrt{3}}\left(\uparrow\downarrow\uparrow\downarrow\uparrow + \uparrow\downarrow\downarrow\uparrow\uparrow \right. \nonumber\\
        &\left. -\downarrow\uparrow\downarrow\uparrow\uparrow - \downarrow\uparrow\uparrow\downarrow\uparrow\right), \nonumber\\ 
    \chi_{9} &= \frac{1}{\sqrt{3}}\left(\uparrow\uparrow\downarrow\downarrow\uparrow + \downarrow\downarrow\uparrow\uparrow\uparrow\right)
        -\frac{1}{2\sqrt{3}}\left(\uparrow\downarrow\uparrow\downarrow\uparrow + \uparrow\downarrow\downarrow\uparrow\uparrow \right. \nonumber\\
        &\left. +\downarrow\uparrow\downarrow\uparrow\uparrow + \downarrow\uparrow\uparrow\downarrow\uparrow\right), \nonumber\\ 
    \chi_{10} &= \frac{1}{2}\left(\uparrow\downarrow\uparrow\downarrow\uparrow - \uparrow\downarrow\downarrow\uparrow\uparrow
        +\downarrow\uparrow\downarrow\uparrow\uparrow -\downarrow\uparrow\uparrow\downarrow\uparrow\right).
    \label{equ:spinbasesexp}
\end{align}

Given the color and spin wavefunctions (\ref{equ:colorbases}), (\ref{equ:spinbases1}), (\ref{equ:spinbases2}) and (\ref{equ:spinbases3}), 
it becomes possible to construct thirty color-spin bases by performing the product of Young tableaux:
\renewcommand{\tabcolsep}{0.1cm}
\renewcommand{\arraystretch}{1}
\begin{equation}
    \begin{aligned}
        \begin{tabular}{|c|}
            \hline
            1 \\
            \cline{1-1}
            2 \\
            \cline{1-1}
            3 \\
            \cline{1-1}
            4 \\
            \cline{1-1}
        \end{tabular}\, \psi_{1}^{\prime},\psi_{1},
        \begin{tabular}{|c|c|c|}
            \hline
            1 & 3 & 4 \\
            \cline{1-3}
            2 \\
            \cline{1-1}
        \end{tabular}\, \psi_{7}^{\prime},\psi_{7},\psi_{13},
        \begin{tabular}{|c|c|}
            \hline
            1 & 4 \\
            \cline{1-2}
            2 \\
            \cline{1-1}
            3 \\
            \cline{1-1}
        \end{tabular}\, \psi_{3}^{\ast},\psi_{3}^{\prime},\psi_{11}^{\prime},\psi_{3},\psi_{11}, \\
        \begin{tabular}{|c|c|}
            \hline
            1 & 3 \\
            \cline{1-2}
            2 & 4 \\
            \cline{1-2}
        \end{tabular}\, \psi_{6}^{\prime},\psi_{6},
        \begin{tabular}{|c|c|c|}
            \hline
            1 & 2 & 4 \\
            \cline{1-3}
            3 \\
            \cline{1-1}
        \end{tabular}\, \psi_{8}^{\prime},\psi_{8},\psi_{14},
        \begin{tabular}{|c|c|}
            \hline
            1 & 3 \\
            \cline{1-2}
            2 \\
            \cline{1-1}
            4 \\
            \cline{1-1}
        \end{tabular}\, \psi_{2}^{\ast},\psi_{4}^{\prime},\psi_{12}^{\prime},\psi_{4},\psi_{12}, \\
        \begin{tabular}{|c|c|}
            \hline
            1 & 2 \\
            \cline{1-2}
            3 & 4 \\
            \cline{1-2}
        \end{tabular}\, \psi_{5}^{\prime},\psi_{5},
        \begin{tabular}{|c|c|c|}
            \hline
            1 & 2 & 3 \\
            \cline{1-3}
            4 \\
            \cline{1-1}
        \end{tabular}\, \psi_{9}^{\prime},\psi_{9},\psi_{15},
        \begin{tabular}{|c|c|}
            \hline
            1 & 2 \\
            \cline{1-2}
            3 \\
            \cline{1-1}
            4 \\
            \cline{1-1}
        \end{tabular}\, \psi_{1}^{\ast},\psi_{2}^{\prime},\psi_{10}^{\prime},\psi_{2},\psi_{10}.
    \end{aligned}
    \label{equ:colorspinbases}
\end{equation}
Due to the Pauli Principle, the wavefunction is inherently fully antisymmetric under the exchange of any pair among the four quarks $q_1$, $q_2$, $q_3$ and $q_4$.
This foundational property allows for the selection of physically possible bases from Eq. (\ref{equ:colorspinbases}) for any pentaquark state 
with a specific flavor configuration and quantum numbers $IJ^P$. These bases are organized and detailed in Table \ref{tab:basis}. 
The explicit expressions of them, as given in Eq. (\ref{equ:colorspinstart})-(\ref{equ:colorspinend}), are employed for the computation of masses and magnetic moments. 
It's noteworthy that these bases are equivalent to that outlined in Ref. \cite{Weng:2019ynv} in numerical calculations, 
which allows the evaluation of eigenvectors for partial width studies.

(1) $J^P=5/2^-$
\begin{equation}
    \psi_{1}^{*} = \phi_{1}\chi_{1}, \quad
    \psi_{2}^{*} = \phi_{2}\chi_{1}, \quad
    \psi_{3}^{*} = \phi_{3}\chi_{1}. \label{equ:colorspinstart}
\end{equation}

(2) $J^P=3/2^-$
\begin{align}
    &\psi_{1}^{\prime} = \frac{1}{\sqrt{3}}\phi_{1}\chi_{5}-\frac{1}{\sqrt{3}}\phi_{2}\chi_{4}+\frac{1}{\sqrt{3}}\phi_{3}\chi_{3}, \\
    &\psi_{2}^{\prime} = -\frac{1}{\sqrt{6}}\phi_{1}\chi_{3}-\frac{1}{\sqrt{3}}\phi_{1}\chi_{4}+\frac{1}{\sqrt{3}}\phi_{2}\chi_{5}
        -\frac{1}{\sqrt{6}}\phi_{3}\chi_{5}, \\
    &\psi_{3}^{\prime} = -\frac{1}{\sqrt{6}}\phi_{1}\chi_{5}+\frac{1}{\sqrt{6}}\phi_{2}\chi_{4}+\sqrt{\frac{2}{3}}\phi_{3}\chi_{3}, \\
    &\psi_{4}^{\prime} = \frac{1}{\sqrt{3}}\phi_{1}\chi_{5}-\frac{1}{\sqrt{6}}\phi_{2}\chi_{3}+\frac{1}{\sqrt{3}}\phi_{2}\chi_{4}
        +\frac{1}{\sqrt{6}}\phi_{3}\chi_{4}, \\
    &\psi_{5}^{\prime} = -\frac{1}{\sqrt{3}}\phi_{1}\chi_{3}+\frac{1}{\sqrt{6}}\phi_{1}\chi_{4}-\frac{1}{\sqrt{6}}\phi_{2}\chi_{5}
        -\frac{1}{\sqrt{3}}\phi_{3}\chi_{5}, \\
    &\psi_{6}^{\prime} = -\frac{1}{\sqrt{6}}\phi_{1}\chi_{5}-\frac{1}{\sqrt{3}}\phi_{2}\chi_{3}-\frac{1}{\sqrt{6}}\phi_{2}\chi_{4}
        +\frac{1}{\sqrt{3}}\phi_{3}\chi_{4}, \\
    &\psi_{7}^{\prime} = -\frac{1}{\sqrt{2}}\phi_{2}\chi_{3}-\frac{1}{\sqrt{2}}\phi_{3}\chi_{4}, \\
    &\psi_{8}^{\prime} = -\frac{1}{\sqrt{2}}\phi_{1}\chi_{3}+\frac{1}{\sqrt{2}}\phi_{3}\chi_{5}, \\
    &\psi_{9}^{\prime} = \frac{1}{\sqrt{2}}\phi_{1}\chi_{4}+\frac{1}{\sqrt{2}}\phi_{2}\chi_{5}, \\
    &\psi_{10}^{\prime} = \phi_{1}\chi_{2}, \\
    &\psi_{11}^{\prime} = \phi_{3}\chi_{2}, \\
    &\psi_{12}^{\prime} = \phi_{2}\chi_{2}.
\end{align}

(3) $J^P=1/2^-$
\begin{align}
    &\psi_{1} = \frac{1}{\sqrt{3}}\phi_{1}\chi_{8}-\frac{1}{\sqrt{3}}\phi_{2}\chi_{7}+\frac{1}{\sqrt{3}}\phi_{3}\chi_{6}, \\
    &\psi_{2} = -\frac{1}{\sqrt{6}}\phi_{1}\chi_{6}-\frac{1}{\sqrt{3}}\phi_{1}\chi_{7}+\frac{1}{\sqrt{3}}\phi_{2}\chi_{8}
        -\frac{1}{\sqrt{6}}\phi_{3}\chi_{8}, \\
    &\psi_{3} = -\frac{1}{\sqrt{6}}\phi_{1}\chi_{8}+\frac{1}{\sqrt{6}}\phi_{2}\chi_{7}+\sqrt{\frac{2}{3}}\phi_{3}\chi_{6}, \\
    &\psi_{4} = \frac{1}{\sqrt{3}}\phi_{1}\chi_{8}-\frac{1}{\sqrt{6}}\phi_{2}\chi_{6}+\frac{1}{\sqrt{3}}\phi_{2}\chi_{7}
        +\frac{1}{\sqrt{6}}\phi_{3}\chi_{7}, \\
    &\psi_{5} = -\frac{1}{\sqrt{3}}\phi_{1}\chi_{6}+\frac{1}{\sqrt{6}}\phi_{1}\chi_{7}-\frac{1}{\sqrt{6}}\phi_{2}\chi_{8}
        -\frac{1}{\sqrt{3}}\phi_{3}\chi_{8}, \\
    &\psi_{6} = -\frac{1}{\sqrt{6}}\phi_{1}\chi_{8}-\frac{1}{\sqrt{3}}\phi_{2}\chi_{6}-\frac{1}{\sqrt{6}}\phi_{2}\chi_{7}
        +\frac{1}{\sqrt{3}}\phi_{3}\chi_{7}, \\
    &\psi_{7} = -\frac{1}{\sqrt{2}}\phi_{2}\chi_{6}-\frac{1}{\sqrt{2}}\phi_{3}\chi_{7}, \\
    &\psi_{8} = -\frac{1}{\sqrt{2}}\phi_{1}\chi_{6}+\frac{1}{\sqrt{2}}\phi_{3}\chi_{8}, \\
    &\psi_{9} = \frac{1}{\sqrt{2}}\phi_{1}\chi_{7}+\frac{1}{\sqrt{2}}\phi_{2}\chi_{8}, \\
    &\psi_{10} = -\frac{1}{2}\phi_{1}\chi_{9}+\frac{1}{2}\phi_{2}\chi_{10}+\frac{1}{\sqrt{2}}\phi_{3}\chi_{10}, \\
    &\psi_{11} = \frac{1}{\sqrt{2}}\phi_{1}\chi_{10}-\frac{1}{\sqrt{2}}\phi_{2}\chi_{9}, \\
    &\psi_{12} = \frac{1}{2}\phi_{1}\chi_{10}+\frac{1}{2}\phi_{2}\chi_{9}-\frac{1}{\sqrt{2}}\phi_{3}\chi_{9}, \\
    &\psi_{13} = -\frac{1}{2}\phi_{1}\chi_{10}-\frac{1}{2}\phi_{2}\chi_{9}-\frac{1}{\sqrt{2}}\phi_{3}\chi_{9}, \\
    &\psi_{14} = \frac{1}{2}\phi_{1}\chi_{9}-\frac{1}{2}\phi_{2}\chi_{10}+\frac{1}{\sqrt{2}}\phi_{3}\chi_{10}, \\
    &\psi_{15} = -\frac{1}{\sqrt{2}}\phi_{1}\chi_{9}-\frac{1}{\sqrt{2}}\phi_{2}\chi_{10}. \label{equ:colorspinend}
\end{align}

\renewcommand{\tabcolsep}{0.5cm} \renewcommand{\arraystretch}{1.5}
\begin{table*}[!htb]
    \caption{Color-spin wave bases of hidden-heavy pentaquarks 
    with isospin $I$ and quantum number $J^{P}$. }
    \label{tab:basis}
    \begin{tabular}{lccc}
    \bottomrule[1.5pt]\bottomrule[0.5pt]
        System &$I$ &$J^{P}$ &Color-spin wave functions \\ \hline
        $nnnQ\bar{Q}$, $sssQ\bar{Q}$ & 3/2
            & ${5/2}^{-}$ & $\psi_{3}^{\ast}$ \\
        &   & ${3/2}^{-}$ & $\psi_{1}^{\prime}$, $\psi_{3}^{\prime}$, $\psi_{11}^{\prime}$ \\
        &   & ${1/2}^{-}$ & $\psi_{1}$, $\psi_{3}$, $\psi_{11}$ \\
        $nnnQ\bar{Q}$ & 1/2
            & ${5/2}^{-}$ & $\frac{1}{\sqrt{2}}\psi_{2}^{\ast}-\frac{1}{\sqrt{2}}\psi_{1}^{\ast}$ \\
        &   & ${3/2}^{-}$ & $\frac{1}{\sqrt{2}}\psi_{4}^{\prime}-\frac{1}{\sqrt{2}}\psi_{2}^{\prime}$,
                $\frac{1}{\sqrt{2}}\psi_{6}^{\prime}-\frac{1}{\sqrt{2}}\psi_{5}^{\prime}$,
                $\frac{1}{\sqrt{2}}\psi_{7}^{\prime}-\frac{1}{\sqrt{2}}\psi_{8}^{\prime}$,
                $\frac{1}{\sqrt{2}}\psi_{12}^{\prime}-\frac{1}{\sqrt{2}}\psi_{10}^{\prime}$ \\
        &   & ${1/2}^{-}$ & $\frac{1}{\sqrt{2}}\psi_{4}-\frac{1}{\sqrt{2}}\psi_{2}$,
                $\frac{1}{\sqrt{2}}\psi_{6}-\frac{1}{\sqrt{2}}\psi_{5}$,
                $\frac{1}{\sqrt{2}}\psi_{7}-\frac{1}{\sqrt{2}}\psi_{8}$,
                $\frac{1}{\sqrt{2}}\psi_{12}-\frac{1}{\sqrt{2}}\psi_{10}$,
                $\frac{1}{\sqrt{2}}\psi_{13}-\frac{1}{\sqrt{2}}\psi_{14}$ \\
        $nnsQ\bar{Q}$, $ssnQ\bar{Q}$ & 1
            & ${5/2}^{-}$ & $\psi_{2}^{\ast}$, $\psi_{3}^{\ast}$ \\
        &   & ${3/2}^{-}$ & $\psi_{1}^{\prime}$, $\psi_{3}^{\prime}$, $\psi_{4}^{\prime}$, $\psi_{6}^{\prime}$,
                $\psi_{7}^{\prime}$, $\psi_{11}^{\prime}$, $\psi_{12}^{\prime}$ \\
        &   & ${1/2}^{-}$ & $\psi_{1}$, $\psi_{3}$, $\psi_{4}$, $\psi_{6}$, $\psi_{7}$, $\psi_{11}$, $\psi_{12}$, $\psi_{13}$ \\
        $nnsQ\bar{Q}$ & 0
            & ${5/2}^{-}$ & $\psi_{1}^{\ast}$ \\
        &   & ${3/2}^{-}$ & $\psi_{2}^{\prime}$, $\psi_{5}^{\prime}$, $\psi_{8}^{\prime}$, $\psi_{9}^{\prime}$, $\psi_{10}^{\prime}$ \\
        &   & ${1/2}^{-}$ & $\psi_{2}$, $\psi_{5}$, $\psi_{8}$, $\psi_{9}$, $\psi_{10}$, $\psi_{14}$, $\psi_{15}$ \\
    \bottomrule[0.5pt]\bottomrule[1.5pt]
    \end{tabular}%
\end{table*}

\section*{Appendix B: Magnetic moments}
\label{apd:moment}
\setcounter{equation}{0}
\renewcommand{\theequation}{B\arabic{equation}}

In this section, we will exhibit the process by which the magnetic moments of pentaquarks are derived. The inspiration for our method is drawn from Eq.(\ref{equ:musum}), 
motivating a computation approach grounded in basic quantum mechanics, where the physical quantity is considered as the average value of an operator.
For hadronic states, the operator $\hat{\mu}$ is applicable solely to the spin components of the wavefunction,
resulting in the neglect of orbital and flavor parts, which are treated as orthogonal. Consequently, in the presence of chromomagnetic mixing, 
it becomes reasonable to utilize the color-spin bases from Table \ref{tab:basis} for the computation of the average value of the operator $\hat{\mu}$.

After finishing the study of masses for given pentaquark states with color-spin bases,
the results yield two types of quantities: the magnetic moments for individual quarks $\mu_i$ and eigenvectors $(C_{1},C_{2},\dots)$.
Considering the example of the system $sssQ\bar{Q}$ with $J=3/2$, the associated color-spin wavefunction is expressed as $\psi=C_{1}\psi_{1}^{\prime}+C_{2}\psi_{3}^{\prime}+C_{3}\psi_{11}^{\prime}$
where the eigenvector is calculated to be $(C_{1},C_{2},C_{3})$. This information allows us to derive the expression for the magnetic moment through the following steps:
\begin{widetext}
\begin{equation}
    \begin{aligned}
        \mu &= \left\langle \psi\left\vert\hat{\mu}\right\vert \psi\right\rangle \\
        &= C_{1}^{2} \left\langle \psi_{1}^{\prime}\left\vert\hat{\mu}\right\vert \psi_{1}^{\prime}\right\rangle
            +C_{2}^{2} \left\langle \psi_{3}^{\prime}\left\vert\hat{\mu}\right\vert \psi_{3}^{\prime}\right\rangle
            +C_{3}^{2} \left\langle \psi_{11}^{\prime}\left\vert\hat{\mu}\right\vert \psi_{11}^{\prime}\right\rangle
            +2C_{1}C_{2} \left\langle \psi_{1}^{\prime}\left\vert\hat{\mu}\right\vert \psi_{3}^{\prime}\right\rangle
            +2C_{1}C_{3} \left\langle \psi_{1}^{\prime}\left\vert\hat{\mu}\right\vert \psi_{11}^{\prime}\right\rangle
            +2C_{2}C_{3} \left\langle \psi_{3}^{\prime}\left\vert\hat{\mu}\right\vert \psi_{11}^{\prime}\right\rangle \\
        &= C_{1}^{2} \left\langle \frac{1}{\sqrt{3}}\phi_{1}\chi_{5}-\frac{1}{\sqrt{3}}\phi_{2}\chi_{4}+\frac{1}{\sqrt{3}}\phi_{3}\chi_{3}\left\vert 
            \hat{\mu}\right\vert \frac{1}{\sqrt{3}}\phi_{1}\chi_{5}-\frac{1}{\sqrt{3}}\phi_{2}\chi_{4}+\frac{1}{\sqrt{3}}\phi_{3}\chi_{3}\right\rangle \\
        &\quad +C_{2}^{2} \left\langle -\frac{1}{\sqrt{6}}\phi_{1}\chi_{5}+\frac{1}{\sqrt{6}}\phi_{2}\chi_{4}+\sqrt{\frac{2}{3}}\phi_{3}\chi_{3}\left\vert 
            \hat{\mu}\right\vert -\frac{1}{\sqrt{6}}\phi_{1}\chi_{5}+\frac{1}{\sqrt{6}}\phi_{2}\chi_{4}+\sqrt{\frac{2}{3}}\phi_{3}\chi_{3}\right\rangle
            +C_{3}^{2} \left\langle \phi_{3}\chi_{2}\left\vert\hat{\mu}\right\vert \phi_{3}\chi_{2}\right\rangle \\
        &\quad +2C_{1}C_{2} \left\langle \frac{1}{\sqrt{3}}\phi_{1}\chi_{5}-\frac{1}{\sqrt{3}}\phi_{2}\chi_{4}+\frac{1}{\sqrt{3}}\phi_{3}\chi_{3}\left\vert 
            \hat{\mu}\right\vert -\frac{1}{\sqrt{6}}\phi_{1}\chi_{5}+\frac{1}{\sqrt{6}}\phi_{2}\chi_{4}+\sqrt{\frac{2}{3}}\phi_{3}\chi_{3}\right\rangle \\
        &\quad +2C_{1}C_{3} \left\langle \frac{1}{\sqrt{3}}\phi_{1}\chi_{5}-\frac{1}{\sqrt{3}}\phi_{2}\chi_{4}+\frac{1}{\sqrt{3}}\phi_{3}\chi_{3}\left\vert 
            \hat{\mu}\right\vert \phi_{3}\chi_{2}\right\rangle
            +2C_{2}C_{3} \left\langle -\frac{1}{\sqrt{6}}\phi_{1}\chi_{5}+\frac{1}{\sqrt{6}}\phi_{2}\chi_{4}+\sqrt{\frac{2}{3}}\phi_{3}\chi_{3}\left\vert 
                \hat{\mu}\right\vert \phi_{3}\chi_{2}\right\rangle \\
        &= \left(\frac{1}{3}C_{1}^{2}+\frac{1}{6}C_{2}^{2}-\frac{\sqrt{2}}{3}C_{1}C_{2} \right)
            \left\langle \chi_{5}\left\vert\hat{\mu}\right\vert \chi_{5}\right\rangle
            +\left(\frac{1}{3}C_{1}^{2}+\frac{1}{6}C_{2}^{2}-\frac{\sqrt{2}}{3}C_{1}C_{2} \right)
            \left\langle \chi_{4}\left\vert\hat{\mu}\right\vert \chi_{4}\right\rangle
            +\left(\frac{1}{3}C_{1}^{2}+\frac{2}{3}C_{2}^{2}+\frac{2\sqrt{2}}{3}C_{1}C_{2} \right)
            \left\langle \chi_{3}\left\vert\hat{\mu}\right\vert \chi_{3}\right\rangle \\
        &\quad +\left(\frac{2}{\sqrt{3}}C_{1}C_{3} + 2\sqrt{\frac{2}{3}}C_{2}C_{3}\right) \left\langle 
            \chi_{3}\left\vert\hat{\mu}\right\vert \chi_{2}\right\rangle
            +C_{3}^{2} \left\langle \chi_{2}\left\vert\hat{\mu}\right\vert \chi_{2}\right\rangle .
    \end{aligned}\label{equ:musteps}
\end{equation}
\end{widetext}
In the last step of Eq.(\ref{equ:musteps}), the color bases are orthogonal and neglected, leaving matrix elements 
$\langle \chi_{i}\left\vert\hat{\mu}\right\vert \chi_{j}\rangle$ in spin space to be determined.

\renewcommand{\tabcolsep}{0.1cm} \renewcommand{\arraystretch}{1.5}
\begin{table*}[!htb]
    \caption{Matrix elements of magnetic moments $\langle \chi_{i}\left\vert\hat{\mu}\right\vert \chi_{j}\rangle$ 
        in spin subspace ($\chi_1$, $\chi_2$, $\chi_3$, $\chi_4$, $\chi_5$).}
    \label{tab:muin12345}
    \begin{tabular}{cccccc}
    \bottomrule[1.5pt]\bottomrule[0.5pt]
        Spin &$\chi_1$ &$\chi_2$ &$\chi_3$ &$\chi_4$ &$\chi_5$ \\ \hline
        $\chi_1$ &$\mu_1+\mu_2+\mu_3+\mu_4+\mu_5$ &0 &0 &0 &0 \\
        $\chi_2$ &0 &$\frac{9}{10}(\mu_1+\mu_2+\mu_3+\mu_4)-\frac{3}{5}\mu_5$ &$-\frac{\sqrt{15}}{30}(\mu_1+\mu_2+\mu_3-3\mu_4)$ &$-\frac{\sqrt{30}}{30}(\mu_1+\mu_2-2\mu_3)$ &$-\frac{\sqrt{10}}{10}(\mu_1-\mu_2)$ \\
        $\chi_3$ &0 &$-\frac{\sqrt{15}}{30}(\mu_1+\mu_2+\mu_3-3\mu_4)$ &$\frac{5}{6}(\mu_1+\mu_2+\mu_3)-\frac{1}{2}\mu_4+\mu_5$ &$-\frac{\sqrt{2}}{6}(\mu_1+\mu_2-2\mu_3)$ &$-\frac{\sqrt{6}}{6}(\mu_1-\mu_2)$ \\
        $\chi_4$ &0 &$-\frac{\sqrt{30}}{30}(\mu_1+\mu_2-2\mu_3)$ &$-\frac{\sqrt{2}}{6}(\mu_1+\mu_2-2\mu_3)$ &$\frac{1}{3}(2\mu_1+2\mu_2-\mu_3)+\mu_4+\mu_5$ &$-\frac{\sqrt{3}}{3}(\mu_1-\mu_2)$ \\
        $\chi_5$ &0 &$-\frac{\sqrt{10}}{10}(\mu_1-\mu_2)$ &$-\frac{\sqrt{6}}{6}(\mu_1-\mu_2)$ &$-\frac{\sqrt{3}}{3}(\mu_1-\mu_2)$ &$\mu_3+\mu_4+\mu_5$ \\
    \bottomrule[0.5pt]\bottomrule[1.5pt]
    \end{tabular}
\end{table*}

\renewcommand{\tabcolsep}{0.1cm} \renewcommand{\arraystretch}{1.5}
\begin{table*}[!htb]
    \caption{Matrix elements of magnetic moments $\langle \chi_{i}\left\vert\hat{\mu}\right\vert \chi_{j}\rangle$ 
        in spin subspace ($\chi_6$, $\chi_7$, $\chi_8$, $\chi_9$, $\chi_{10}$).}
    \label{tab:muin678910}
    \begin{tabular}{cccccc}
    \bottomrule[1.5pt]\bottomrule[0.5pt]
        Spin &$\chi_6$ &$\chi_7$ &$\chi_8$ &$\chi_9$ &$\chi_{10}$ \\ \hline
        $\chi_6$ &$\frac{1}{9}(5\mu_1+5\mu_2+5\mu_3-3\mu_4-3\mu_5)$ &$-\frac{\sqrt{2}}{9}(\mu_1+\mu_2-2\mu_3)$ &$-\frac{\sqrt{6}}{9}(\mu_1-\mu_2)$ &$-\frac{\sqrt{6}}{9}(\mu_1+\mu_2-2\mu_3)$ &$-\frac{\sqrt{2}}{3}(\mu_1-\mu_2)$ \\
        $\chi_7$ &$-\frac{\sqrt{2}}{9}(\mu_1+\mu_2-2\mu_3)$ &$\frac{1}{9}(4\mu_1+4\mu_2-2\mu_3+6\mu_4-3\mu_5)$ &$-\frac{2\sqrt{3}}{9}(\mu_1-\mu_2)$ &$-\frac{\sqrt{3}}{9}(2\mu_1+2\mu_2-\mu_3-3\mu_4)$ &$\frac{1}{3}(\mu_1-\mu_2)$ \\
        $\chi_8$ &$-\frac{\sqrt{6}}{9}(\mu_1-\mu_2)$ &$-\frac{2\sqrt{3}}{9}(\mu_1-\mu_2)$ &$\frac{1}{3}(2\mu_3+2\mu_4-\mu_5)$ &$\frac{1}{3}(\mu_1-\mu_2)$ &$-\frac{\sqrt{3}}{3}(\mu_3-\mu_4)$ \\
        $\chi_9$ &$-\frac{\sqrt{6}}{9}(\mu_1+\mu_2-2\mu_3)$ &$-\frac{\sqrt{3}}{9}(2\mu_1+2\mu_2-\mu_3-3\mu_4)$ &$\frac{1}{3}(\mu_1-\mu_2)$ &$\mu_5$ &0 \\
        $\chi_{10}$ &$-\frac{\sqrt{2}}{3}(\mu_1-\mu_2)$ &$\frac{1}{3}(\mu_1-\mu_2)$ &$-\frac{\sqrt{3}}{3}(\mu_3-\mu_4)$ &0 &$\mu_5$ \\
    \bottomrule[0.5pt]\bottomrule[1.5pt]
    \end{tabular}
\end{table*}

In Tables \ref{tab:muin12345} and \ref{tab:muin678910}, the matrix elements of magnetic moment in spin subspaces
($\chi_1$, $\chi_2$, $\chi_3$, $\chi_4$, $\chi_5$) and ($\chi_6$, $\chi_7$, $\chi_8$, $\chi_9$, $\chi_{10}$), respectively, are provided directly.
These matrix elements are determined with the assistance of the spin bases given in Eq. (\ref{equ:spinbasesexp}).
The expressions involve the values $\mu_i$, where the indexes $i$ represent flavors in the configuration $q_1q_2q_3q_4\bar{q}_5$,
and the $\mu_i$ have been calculated previously for individual quarks.

It is important to note that the element is reasonable only if $\chi_{i}$ and $\chi_{j}$ belong to the same spin multiplets,
as this is the realm where chromomagnetic mixing occurs. Going forward, these methods can be consistently applied to evaluate magnetic moments for pentaquarks. 
The availability of color-spin bases, alongside the calculated eigenvectors and $\mu_i$, facilitates this systematic approach.

\end{document}